\documentclass[10pt]{article}
\pdfoutput=1
\usepackage[T1]{fontenc}
\usepackage{amssymb,amsmath,mathrsfs,enumerate}
\usepackage{graphicx,rotate,multicol}
\usepackage{wrapfig}
\usepackage[colorlinks=true,
            linkcolor=red,
            urlcolor=blue,
            citecolor=blue]{hyperref}
\usepackage{color}
\usepackage{soul}
\usepackage{cite}
\usepackage{verbatim}
\long\def\rpl#1!!#2!!{\textcolor{red}{#1} \textcolor{blue}{#2}}

\usepackage{cleveref}
\usepackage[makeroom]{cancel}
\usepackage[normalem]{ulem}
\usepackage{amsmath,amssymb}
\usepackage{slashed}
\usepackage{xcolor}
\usepackage{graphicx}
\usepackage{cite}
\usepackage{pdflscape}
\usepackage{multirow}
\usepackage[thinlines]{easytable}
\usepackage{url}
\usepackage[utf8]{inputenc}
\usepackage{longtable}
\usepackage{booktabs}
\usepackage{graphicx,floatflt,amssymb,rotate, cancel}
\usepackage{mathrsfs}
\usepackage{amssymb}
\usepackage{amsmath}
\usepackage{subcaption}
\usepackage{float}
\usepackage{pstricks}
\usepackage[toc,page]{appendix}
\usepackage[mathscr]{euscript}
\usepackage{bbold}
\usepackage{color}

 \newcommand{\gsim}{\raisebox{-0.13cm}{~\shortstack{$>$ \\[-0.07cm]
       $\sim$}}~}
 \newcommand{\lsim}{\raisebox{-0.13cm}{~\shortstack{$<$ \\[-0.07cm]
       $\sim$}}~}

\def \order(#1){{\cal O} \left(#1 \right)}

\evensidemargin=\oddsidemargin
\allowdisplaybreaks

\DeclareUnicodeCharacter{2212}{-}
\begin{document}

\begin{flushright}
HRI-RECAPP-2021-002
\end{flushright}

\begin{center}
	{\Large \bf Generalized 2HDM with wrong-sign lepton-Yukawa coupling, in light of $g_{\mu}-2$ and lepton flavor violation at the future LHC } \\
	\vspace*{1cm} {\sf Nivedita
          Ghosh$^{a,}$\footnote{niveditaghosh@hri.res.in},~Jayita
          Lahiri$^{b,}$\footnote{jayitalahiri@rnd.iitg.ac.in}} \\
	\vspace{10pt} {\small \em 
          $^a$Regional Centre for Accelerator-based Particle Physics,
Harish-Chandra Research Institute, HBNI,
Chhatnag Road, Jhunsi, Allahabad - 211 019, India. \\
          $^b$Department of Physics, Indian Institute of Technology Guwahati,
North Guwahati, Assam - 781039, India.}
	
	\normalsize
\end{center}

%
%
%
%
%
%
%
%


\begin{abstract}

 To explain the observed muon anomaly and simultaneously evade bounds from lepton flavor violation in the same model parameter space is a long-cherished dream. In view of a generalized Two Higgs Doublet Model, with a Yukawa structure as a perturbation of Type-X, we are able to get substantial parameter space satisfying these criteria. In this work, we focus on a region with ``{\bf wrong-sign}'' lepton-Yukawa coupling which gives rise to interesting phenomenological consequences. Performing a simple cut-based analysis, we show that at 14 TeV run of the LHC with $300 fb^{-1}$ integrated luminosity, part of the model parameter space can be probed with significance $\gsim 5\sigma$ which further improves with Artificial Neural Network analysis.

\end{abstract}

\bigskip

\section{Introduction}

The discovery of the 125-GeV scalar at the LHC~\cite{Aad:2012tfa,Chatrchyan:2012xdj} with its close resemblance to the Standard Model (SM) Higgs boson puts a stringent limit on the New Physics (NP) scenarios. However, at the same time, various experimental evidence have convinced us by now that the SM is not the complete theory. The anomalous magnetic moment of muon is one such observation which urges the physicists to go beyond the SM. There was a long-standing discrepancy of $\sim 3.7 \sigma$ between the SM prediction and the experimental observation~\cite{Blum:2013xva} which has increased to $\sim 4.2\sigma$ deviation with increasing precision, as reported by ``MUON G-2'' collaboration at the Fermilab~\cite{Grange:2015fou} in their first run of data. The future J-PARC experiment~\cite{Iinuma:2011zz}  will help us achieve better understanding of it in the future.  

On the contrary, the Lepton Flavor Violation (LFV) has not been observed in the charged lepton sector, although it has been confirmed in the neutrino sector years ago in the neutrino oscillation experiments~\cite{Fukuda:1998mi,Ahmad:2002jz}. However, various low energy experiments~\cite{Bellgardt:1988qe,Wintz:1998rp,Kuno:1999jp,Aubert:2009ag,TheMEG:2016wtm,Lindner:2016bgg,Bartolotta:2017mff,Beringer:1900zz,Endo:2020mev,Iguro:2020rby} have been able to put strong upper limits on the branching ratios of LFV decays of charged leptons. 

These two phenomena, namely, the muon anomalous magnetic moment and lepton flavor violation are not independent. The models which predict LFV will have severe constraints from the observation of muon anomalous magnetic moment~\cite{Lindner:2016bgg,Li:2019xmi}. Typically the models which can explain muon anomalous magnetic moment will predict the masses of the heavy states running in the loop at a lower range which may be in tension with the non-observation of LFV. Therefore in the context of models which predict lepton flavor violation and can explain muon anomaly, it is extremely important to answer questions such as: 1) Is it possible to explain muon anomalous magnetic moment in some regions of the parameter space while obeying LFV constraints at the same time? 2) Is simultaneous observation of muon anomalous magnetic moment and lepton flavor violating processes in the respective experiments possible? 3) Moreover, is it possible to look for LFV at the collider experiments which will be a complementary approach to the low energy experiments. There has been considerable work in this direction in the past in the context of 2HDM~\cite{Atwood:1996vj,DiazRodolfo:2000yy,Diaz:2000cm,Xiao:2003ya,Cao:2009as,Arhrib:2011wc,Broggio:2014mna,Wang:2014sda,Ilisie:2015tra,Abe:2015oca,Han:2015yys,Liu:2015oaa,Chun:2015hsa,Omura:2015xcg,Omura:2015nja,Davidson:2016utf,Banerjee:2016foh,Primulando:2016eod,Chun:2016hzs,Cherchiglia:2016eui,Arhrib:2017yby,Cherchiglia:2017uwv,Wang:2018hnw,Chakrabarty:2018qtt,Iguro:2019sly,Chun:2019oix,Primulando:2019ydt,Jana:2020pxx,Frank:2020smf,Chun:2020uzw,Abdallah:2020vgg,Rose:2020nxm,Capdevilla:2021rwo,Yin:2020afe,Jana:2020joi,Li:2020dbg}. We mention here that in all of the existing literature mentioned above, the muon anomalous magnetic moment, the prospect of lepton flavor violation and the collider search for LFV decays have been considered individually.  

We addressed all these questions simultaneously in the context of generalized 
Two Higgs Doublet Model(2HDM)~\cite{Mahmoudi:2009zx,DiazCruz:2010yq,Bai:2012ex,Jueid:2021avn}, with a Yukawa structure as a perturbation of Type X 2HDM, where we have shown that the presence of non-standard light scalars allows one to satisfy both muon anomaly and LFV constraints in specific regions of the parameter space. We focus on the ``wrong-sign'' region, which is an experimentally viable scenario~\cite{Ferreira:2014naa,Han:2020zqg} and leads to interesting phenomenology.

After satisfying both muon anomaly and LFV constraints at two-loop, we impose all relevant theoretical and experimental constraints on the model parameter space. Furthermore, we perform a collider analysis in $\ell^{+}\ell'^{-} + \slashed{E_T}$ channel, where $\ell, \ell'= e,\mu$. These final states result from the flavor-violating decay of the CP-odd scalar, $A \to \ell \tau_{\ell'}$, where $\tau_{\ell'}$ implies $\tau$ decaying leptonically. 
With simple cut-based analysis we show that in the ``wrong-sign'' case, an interesting region of parameter space can be probed in the 14 TeV LHC. Notably, larger parameter space can be probed in this case, with lower luminosity compared to the ``right-sign'' case~\cite{Ghosh:2020tfq}. We then perform an Artificial Neural Network(ANN) analysis and observe significant improvement over our cut-based results.

The paper is organized as follows. In section~\ref{model} we briefly discuss the model considered in this work and describe the ``wrong-sign'' region of it. Having discussed the muon anomaly and its impact on our model parameter space in section~\ref{muonanomaly}, we move to section~\ref{constraints}, where we explore the allowed parameter space taking into account the bounds from low energy observables, theoretical and experimental constraints.
We present a cut-based as well as neural-network-based collider analysis in section~\ref{collider}. We summarize our results and conclude in section~\ref{conclusion}.


\section{The ``wrong-sign'' region of the Model}\label{model}

In this work, we have considered the generalized two Higgs doublet model with the Yukawa structure as a perturbation from Type-X 2HDM. This specific choice for the Yukawa structure is motivated by the observed $(g-2)_{\mu}$ data while at the same time we want to probe the lepton flavor violation in the extended scalar sector~\cite{Mahmoudi:2009zx,DiazCruz:2010yq,Bai:2012ex}. We follow the convention as in~\cite{Primulando:2016eod}~\footnote{For general 2HDM review 
one should look into Ref~\cite{Branco:2011iw}.}. Two complex scalar doublets $\Phi_{1}$ and $\Phi_{2}$ with hypercharge $Y=1$~\footnote{We abide by the convention $Q=T_3+\frac{Y}{2}$} are present in this model leading to the most general scalar 
potential as follows:
\begin{equation}
\begin{aligned}
	{\cal V}_{2HDM} &= m_{11}^2 (\Phi_1^{\dagger} \Phi_1^{\phantom{\dagger}}) + m_{22}^2 (\Phi_2^{\dagger} \Phi_2^{\phantom{\dagger}}) -  [m_{12}^2(\Phi_1^{\dagger} \Phi_2^{\phantom{\dagger}}) + \text{h.c.}]  \\
&+ \frac{1}{2} \lambda_1 (\Phi_1^{\dagger} \Phi_1^{\phantom{\dagger}})^2 + \frac{1}{2} \lambda_2 (\Phi_2^{\dagger} \Phi_2^{\phantom{\dagger}})^2 + \lambda_3 (\Phi_1^{\dagger} \Phi_1^{\phantom{\dagger}})(\Phi_2^{\dagger} \Phi_2^{\phantom{\dagger}}) + \lambda_4 (\Phi_1^{\dagger} \Phi_2^{\phantom{\dagger}})(\Phi_2^{\dagger} \Phi_1^{\phantom{\dagger}}) \\
&+ \{\frac{1}{2} \lambda_5 (\Phi_1^{\dagger} \Phi_2^{\phantom{\dagger}})^2 + [\lambda_6 (\Phi_1^{\dagger} \Phi_1^{\phantom{\dagger}}) + \lambda_7 (\Phi_2^{\dagger} \Phi_2^{\phantom{\dagger}})] (\Phi_1^{\dagger} \Phi_2^{\phantom{\dagger}}) + \text{h.c.} \}.
\end{aligned}
\end{equation}
where ${h.c.}$ denotes the $\rm{Hermitian~Conjugate}$ term.

We have assumed CP is conserved in the Higgs sector, therefore $m_{12}^2$, $\lambda_5$, $\lambda_6$ and $\lambda_7$ are taken to be real along with all the other parameters. Moreover, in the absence of $Z_2$ symmetry ($\Phi_1 \rightarrow \Phi_1, \Phi_2 \rightarrow - \Phi_2$) $\lambda_6$ and $\lambda_7$ are taken to be non-zero.
 Diagonalizing the mass matrix for the CP-even neutral states we get the mass eigenstates $h$ and $H$. In principle, either $h$ or $H$ can behave like the Higgs of Standard Model with mass 125 GeV, which is the so-called ``alignment limit".

Having briefly discussed the Higgs potential of our model, we proceed towards the Yukawa sector. We focus here on the so-called ``wrong-sign'' region of the Yukawa Lagrangian. We will see that this region gives rise to interesting and unique phenomenological consequences which will be very different from the ``right-sign'' regime which has been considered in detail in~\cite{Ghosh:2020tfq}. 
In the generalized 2HDM, no $Z_2$ symmetry is imposed on the Yukawa Lagrangian, and therefore this model generates tree-level flavor changing neutral current(FCNC), a phenomenon which is of primary interest to us. In this case, the Yukawa Lagrangian takes the most general form: 

\begin{equation}
-{\cal L}_{Yukawa} = \bar{Q}_L (Y^d_1\Phi_1 + Y^d_2\Phi_2) d_R + \bar{Q}_L (Y^u_1 \tilde{\Phi}_1 + Y^u_2 \tilde{\Phi}_2) u_R + \bar{L}_L (Y^\ell_1\Phi_1 + Y^\ell_2\Phi_2) e_R + h.c.
\label{type3yuk}
\end{equation}

In Eq.~\ref{type3yuk}$, Y_{1,2}^{u,d,\ell}$ are the Yukawa matrices whose flavor indices have been suppressed and $\tilde{\Phi}_i = i\sigma_2\Phi_i^*$. 
Without assuming any particular relation between the matrices $Y_1$ and $Y_2$ it is impossible to diagonalize the two of them simultaneously, which leads to tree-level scalar mediated FCNC. As we consider the Yukawa Lagrangian as a perturbation of Type X model~\cite{Crivellin:2015hha} in terms of FCNC couplings, we diagonalize $Y_2^u$, $Y_2^d$ and $Y_1^\ell$ matrices whereas $Y_1^u$, $Y_1^d$ and $Y_2^\ell$ remain non-diagonal resulting in LFV. After diagonalization, the Yukawa Lagrangian involving the neutral scalars takes the following form.

 \begin{align}
 -{\cal L}^\phi_{Yukawa} & = \bar u_L \left[ \left( \frac{c_\alpha {\bf m}^u}{v s_\beta} - \frac{c_{\beta-\alpha} \Sigma^u}{\sqrt{2} s_\beta}\right) h + \left( \frac{s_\alpha {\bf m}^u}{ s_\beta v} + \frac{s_{\beta-\alpha} \Sigma^u}{\sqrt{2} s_\beta}\right) H \right] u_R \nonumber \\
&+ \bar d_L \left[ \left( \frac{c_\alpha {\bf m}^d}{v s_\beta} - \frac{c_{\beta-\alpha} \Sigma^d}{\sqrt{2} s_\beta}\right) h + \left( \frac{s_\alpha {\bf m}^d}{ s_\beta v} + \frac{s_{\beta-\alpha} \Sigma^d}{\sqrt{2} s_\beta}\right) H \right] d_R \nonumber \\
 &+ \bar e_L \left[ \left( - \frac{s_\alpha {\bf m}^\ell}{v c_\beta} + \frac{c_{\beta-\alpha} \Sigma^\ell}{\sqrt{2} c_\beta}\right)h + \left( \frac{c_\alpha {\bf m}^\ell}{ c_\beta v} - \frac{s_{\beta-\alpha} \Sigma^\ell}{\sqrt{2} c_\beta}\right) H \right] e_R \nonumber \\
 %
 %
& - i \left[ \bar u_L \left( \frac{ {\bf m}^u}{ t_\beta v} - \frac{ \Sigma^u }{\sqrt{2} s_\beta}\right) u_R 
 + \bar d_L \left(- \frac{ {\bf m}^d}{ t_\beta v} + \frac{ \Sigma^d }{\sqrt{2} s_\beta}\right) d_R 
 + \bar e_L \left( \frac{ t_\beta{\bf m}^\ell}{ v} - \frac{ \Sigma^\ell }{\sqrt{2} c_\beta}\right) e_R 
 \right] A + h.c.  \label{eq:Yu_phi}
 \end{align}

Here ${\bf m}^f$ the diagonal mass matrices of the fermions, $\Sigma^u = U_L^u Y_1^u U_R^{{\dagger}u}$, $\Sigma^d = U_L^d Y_1^d U_R^{{\dagger}d}$ and $\Sigma^\ell = U_L^\ell Y_2^u U_R^{{\dagger}l}$. 
$c_\alpha = \cos \alpha, s_\alpha = \sin \alpha, c_{\beta - \alpha} = \cos(\beta - \alpha), s_{\beta - \alpha} = \sin(\beta - \alpha) $ and $t_{\beta} = \tan \beta$, where $U_L$ and $U_R$ are the unitary matrices which diagonalize the Yukawa matrices and the angle $\tan \beta$ is the ratio of the VEVs of the two doublets $v_1$ and $v_2$ and $\alpha$ is the mixing angle between the neutral CP-even components of the two doublets. The flavor-changing vertices are the effects of non-zero $\Sigma^{f}$ matrices. Notably, the non-diagonal couplings of the pseudoscalar $A$ (see Eq.~\ref{eq:Yu_phi}) play the most important role in our study and we will call them $y_{\mu e}$, $y_{\tau e}$ and $y_{\mu\tau}$ henceforth.

The Yukawa couplings of the charged Higgs boson ($H^{\pm}$) can be written as

\begin{align}
{\cal L}^{H^\pm}_Y & = \frac{\sqrt{2}}{v} \bar u_{i }  \left( m^u_i  \xi^{u*}_{ki}  V_{kj} P_L + V_{ik}  \xi^d_{kj}  m^d_j P_R \right) d_{j}  H^+  
+ \frac{\sqrt{2}}{v}  \bar \nu_i  \xi^\ell_{ij} m^\ell_j P_R \ell_j H^+ + h.c. \label{eq:Yukawa_CH}
\end{align}

Here $V\equiv U^u_L U^{d\dagger}_L$ is the Cabibbo-Kobayashi-Maskawa (CKM) matrix, $P_{R,L}=(1 \pm \gamma_5)/2$ and $\xi^{f}$ matrices are defined as the following.

\begin{eqnarray}
\xi^{u}_{ij} = \frac{1}{t_{\beta}} \delta_{ij} - \frac{v}{\sqrt{2}s_{\beta}} \frac{{\Sigma^u}_{ij}}{m^u_j}, \\
\xi^{d}_{ij} = -\frac{1}{t_{\beta}} \delta_{ij} + \frac{v}{\sqrt{2}s_{\beta}} \frac{{\Sigma^d}_{ij}}{m^d_j}, \\
\xi^{\ell}_{ij} = t_{\beta} \delta_{ij} - \frac{1}{\sqrt{2}c_{\beta}} \frac{{\Sigma^\ell}_{ij}}{m^\ell_j}
\end{eqnarray}

Having discussed the general Yukawa structure of this model we proceed to the ``wrong-sign'' region which is the focus of this work. In principle, the 125-GeV Higgs can have the SM-like coupling (``right-sign'') as well as the ``wrong-sign'' Yukawa coupling. The generic conditions for right- and wrong-sign Yukawa coupling of the fermions are as follows. 
\begin{eqnarray}
&&y_{h_{SM}}^{f_i}~\times~y^{V}_{h_{SM}} > 0~{\rm for~``right-sign"},\label{rightsign}\\
&&y_{h_{SM}}^{f_i}~\times~y^{V}_{h_{SM}} < 0~{\rm for~``wrong-sign"}.\label{wrongsign}
\end{eqnarray}

Where $y^{V}_{h_{SM}}$ denotes the coupling of the SM-like Higgs to the vector bosons and $y_{h_{SM}}^{f_i}$ are the SM-Higgs coupling to the fermions (up-type quarks, down-type quarks, and leptons) respectively, normalized to their SM values. However, it is to be noted that the wrong-sign case in the up-type quark sector is disfavored from the $h_{SM} \rightarrow \gamma\gamma$ data~\cite{Ferreira:2014naa}, whereas in the down-type and lepton sector wrong-sign is phenomenologically viable~\cite{Ferreira:2014naa,Han:2020zqg}. In our analysis, we are interested in the wrong-sign Yukawa coupling in the lepton sector.
The absolute values of $y_{h_{SM}}^{\ell}$ and $y^{V}_{h_{SM}}$ have to be close to unity because of the restrictions of 125-GeV Higgs signal strength data~\cite{Sirunyan:2018koj,Aad:2019mbh}. In the ``wrong-sign'' regime, the two couplings are mutually opposite in sign.
Moreover, there are two possible scenarios as mentioned earlier, (a) The lightest CP-even scalar $h$ is SM-like ie. $m_h = m_{h_{SM}} = 125$ GeV and (b) when the heavier CP-even scalar $H$ is SM-like, ie. $m_H = m_{h_{SM}} = 125$ GeV. Both the scenarios can correspond to ``wrong-sign'' lepton-Yukawa coupling depending on the conditions stated in Eq.~\ref{wrongsign}.

First, let us consider the first scenario in the ``wrong-sign'' region. Here the 125-GeV Higgs couplings are given by the following relations:
\begin{equation}
y^{V}_h\simeq \sin(\beta-\alpha) 
\end{equation}
\begin{equation}  
y_h^{\ell}=\sin(\beta - \alpha) - \cos(\beta -\alpha)\tan \beta + \frac{\cos(\beta-\alpha) \Sigma^\ell}{\sqrt{2} \cos\beta}
\label{yukawa1}
\end{equation}
The gauge boson couplings of 125-GeV Higgs are experimentally found to be close to their SM predictions and therefore it is ideal to assume $|\sin(\beta - \alpha)| \approx 1$ and $|\cos(\beta - \alpha)| << 1$.
When $\sin(\beta-\alpha) > 0$ and $\cos(\beta-\alpha) > 0$ and $\tan \beta$ is large($\gsim 10$), $y_h^{\ell}$ becomes negative and the product $y_h^{\ell}~\times~y^{V}_h < 0$. This scenario corresponds to the ``wrong-sign'' lepton-Yukawa coupling. 
In this limit, $y_h^{\ell}$ takes the form of $-(1\pm\epsilon)$ where $\epsilon$ is a very small positive quantity.
One should note, it follows directly from Eq.~\ref{yukawa1} that the last term in the coupling $y_h^{\ell}$ i.e. $\frac{\cos(\beta-\alpha) \Sigma^\ell}{\sqrt{2} \cos\beta}$ has negligible contribution. The reason behind this are as follows. In our model, ie. the perturbative limit of Type-X 2HDM, the diagonal elements of $\Sigma^\ell$ matrices are assumed to be small, justified by the fact that the 125-GeV scalar couplings to the leptons in Eq.~\ref{eq:Yu_phi} or \ref{yukawa1} are expected to be mostly SM-like. In addition, in the alignment limit, $|\cos(\beta-\alpha)| << 1$ causing a further suppression.

Next we consider the ``wrong-sign'' region in the second scenario ie. when the heavier CP-even Higgs $H$ is SM-like. In this case, the following relations hold.
\begin{equation}
y^{V}_H\simeq \cos(\beta-\alpha)   
\end{equation}
\begin{equation}  
y_H^{\ell}=\cos(\beta - \alpha) + \sin(\beta -\alpha)\tan \beta - \frac{\sin(\beta-\alpha) \Sigma^\ell}{\sqrt{2} \cos\beta}
\label{yukawa2}
\end{equation}
Here too, the gauge boson couplings are assumed to be in the SM-ballpark and consequently $|\cos(\alpha - \beta)| \approx 1$ and $|\sin(\beta - \alpha)| << 1$.
When $\sin(\beta-\alpha) < 0$ and $\cos(\beta-\alpha) > 0$ and $\tan \beta$ is large($\gsim 10$), $y_H^{\ell}$ becomes negative and the product $y_H^{\ell}~\times~y^{V}_H < 0$. This scenario corresponds to the ``wrong-sign'' lepton-Yukawa coupling. The last term in Eq.~\ref{yukawa2} ie. $\frac{\sin(\beta-\alpha) \Sigma^\ell}{\sqrt{2} \cos\beta}$ makes tiny contribution to the lepton-Yukawa couplings of the SM-like Higgs, when $|\sin(\beta - \alpha)| << 1$ and the diagonal elements of $\Sigma^{\ell}$ matrix are close to 0, like the previous scenario. Here, $y_H^{\ell}$ takes the form of $-(1\pm\epsilon)$, $\epsilon$ being a very small positive quantity. In the present work, we will focus on the first scenario ie. the lightest CP-even scalar is the SM-like Higgs, for reasons which will be discussed in detail shortly.


\section{Anomalous magnetic moment of muon}
\label{muonanomaly}

The experimentally observed muon anomalous magnetic moment is an impressively precise measurement which helps us probe the higher-order quantum corrections to a large degree of precision. Moreover, it also indicates the existence of new physics because of the long-standing discrepancy between SM prediction and experimental observation~\cite{Blum:2013xva}. 

The tree level the value of $g_{\mu}$(gyromagnetic ratio for $\mu$) is 2. It receives correction from loop effects parameterized in terms of $a_{\mu} = \frac{g_{\mu}-2}{2}$ in quantum field theory. In the SM, it receives contribution via QED, electroweak and hadronic loops. The SM contributions up to three orders in the electromagnetic constant, has been calculated by~\cite{Davier:2010nc,Hagiwara:2011af,Davier:2017zfy,Davier:2019can}. Taking into account pure QED, electroweak and hadronic contribution, the SM prediction for muon anomaly has been calculated~\cite{Aoyama:2020ynm,Davier:2017zfy,Keshavarzi:2018mgv,Colangelo:2018mtw,Hoferichter:2019mqg,Davier:2019can,Keshavarzi:2019abf,Kurz:2014wya,Melnikov:2003xd,Masjuan:2017tvw,Colangelo:2017fiz,Hoferichter:2018kwz,Gerardin:2019vio,Bijnens:2019ghy,Colangelo:2019uex,Colangelo:2014qya,Blum:2019ugy,Aoyama:2012wk,Czarnecki:2002nt,Aoyama:2019ryr,Gnendiger:2013pva,Zyla:2020zbs}. The most recent estimate being~\cite{Aoyama:2020ynm}

\begin{equation}
a_{\mu}^{SM}= 116591810(43)  \times 10^{-11}
\label{eq:sm}
\end{equation}

\noindent
Recently, the ``MUON G-2'' collaboration at Fermilab~\cite{Grange:2015fou} has published their result~\cite{Albahri:2021ixb}.
\begin{equation}
a_{\mu}^{exp-FNAL} = 116592040(54)  \times 10^{-11}
\end{equation}
The combined new world average(combination of recent FNAL~\cite{Albahri:2021ixb} and older BNL(2006)~\cite{Bennett:2006fi} data) is published as~\cite{Abi:2021gix}
\begin{equation}
a_{\mu}^{exp-comb} = 116592061(41)  \times 10^{-11}
\label{eq:exp}
\end{equation}

\noindent
The difference between the experimental observation and the SM prediction, defined as $\Delta a_{\mu}$, amounts to a $4.2\sigma$ discrepancy, which urges us to look beyond the SM.
\begin{equation}
 \Delta a_{\mu} = a_{\mu}^{exp-comb} - a_{\mu}^{SM} = 251(59) \times 10^{-11}
 \label{mgm2}
\end{equation}

\noindent
One can compare this new result with the earlier BNL(2006) result~\cite{Bennett:2006fi}. 

\begin{equation}
a_{\mu}^{exp-BNL} = 116592089(63)  \times 10^{-11}
\end{equation}

\noindent
Earlier the difference between the SM prediction and experiment resulted in a $3.7\sigma$ discrepancy.
\begin{equation}
\Delta a_{\mu} ^{BNL}=  a_{\mu}^{exp-BNL} - a_{\mu}^{SM} = 279(76)  \times 10^{-11}
\label{mgm2_old}
\end{equation}

\noindent
We have considered one loop as well as two-loop Bar-Zee type contribution~\cite{Ghosh:2020tfq} to $\Delta a_{\mu}$ in generalized 2HDM, within the framework of ``wrong-sign'' region of lepton-Yukawa coupling. The major contribution comes from two-loop Bar-Zee diagrams involving heavy fermions such as $t, b, \tau$ running in the loop. Since in our case, the lepton coupling to the pseudo(scalar) is enhanced, the $\tau$ loop gives the most dominant contribution. These two-loop contributions exceed the one-loop contribution and have been studied in earlier works~\cite{Queiroz:2014zfa,Ilisie:2015tra}. It has been pointed out that, despite having a loop suppression factor, the two-loop diagrams receive an enhancement factor of $\frac{M^2}{m_{\mu}^2}$, where $M$ is the mass of heavy fermion in the loop.
We have also considered all other Bar-Zee diagrams which make sub-dominant contributions in general but can be important in some regions of the parameter space~\cite{Ilisie:2015tra}.  

\begin{figure}[!hptb]
\centering
\includegraphics[width=70mm,height=60.0mm]{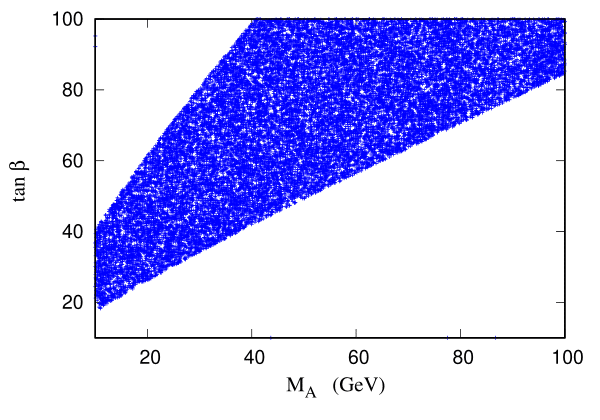}
\caption{\it The allowed region in $m_A - \tan \beta$ plane from $g_{\mu}-2$ data at $3 \sigma$. The flavor changing couplings are taken to be $y_{\mu e}= 10^{-7}, y_{\tau e}=4\times 10^{-5}, y_{\mu \tau}=4\times 10^{-5}$. The non-standard neutral CP-even Higgs mass is 450 GeV and charged Higgs mass is 460 GeV.}
\label{muon_anomaly}
\end{figure}

We compute $\Delta a_{\mu}$ taking into account all the one and two-loop contributions following~\cite{Queiroz:2014zfa,Ilisie:2015tra}. We scan the parameter space of our model in the ``wrong-sign'' Yukawa region and plot the allowed region in the $m_A - \tan \beta$ plane in Fig.~\ref{muon_anomaly}. For the scanning, the flavor changing couplings are taken to be $y_{\mu e}= 10^{-7}, y_{\tau e}=4\times 10^{-5}, y_{\mu \tau}=4\times 10^{-5}$. The non-standard neutral CP-even Higgs mass and charged Higgs mass are fixed at 450 GeV and 460 GeV. The choice of non-standard scalar masses will be justified in the next sections. We mention here that, compared to earlier works in the context of $(g-2)_{\mu}$ in Type X 2HDM~\cite{Broggio:2014mna,Cherchiglia:2017uwv}, we have considered the most updated experimental bound~\cite{Albahri:2021ixb,Abi:2021gix}, exhaustive set of one- and two-loop diagrams and also the effect of lepton flavor violating vertices. We mention here that, the contribution of the lepton flavor violating vertices to this calculation is negligible owing to the smallness of the lepton flavor violating couplings.

Low mass pseudoscalar with an enhanced coupling to the $\tau$ leptons will give a significant contribution to $\Delta a_{\mu}$. In our model, the coupling of pseudoscalar with a pair of $\tau$ leptons is proportional to $\tan \beta$. Therefore low $m_A$ and large $\tan\beta$ region is favored in the light of $g_{\mu}-2$ data. While scanning the parameter space we have used the $3\sigma$ bound on the experimentally observed central value of $\Delta a_{\mu}$(Eq.~\ref{mgm2}). Although the pseudoscalar couplings being proportional to $\tan \beta$ do not depend on the right- or wrong-sign, the couplings of the CP-even scalar($H$) do depend on that. In the ``wrong-sign'' region the contribution from the CP-even scalar interferes destructively with that of the pseudoscalar, while in the ``right-sign'' case the interference is constructive.
In our case, the masses of the CP-even scalar ($m_H$) and charged scalar ($m_{H^{\pm}}$) are taken to be much larger compared to the pseudoscalar mass($m_A$), a choice which we will justify shortly. Because of larger masses those scalars contribute minimally compared to the light pseudoscalar and the aforementioned interference is insignificant. However, one should note that, if masses of the non-standard CP-even scalar masses are comparable with the pseudoscalar mass, such interference will play an important role in defining the allowed contour and in the wrong-sign case, different regions of parameter space compared to the right-sign case, may open up.

\section{Constraints on the model}\label{constraints}

From our discussion in the previous section, it is clear that the major contribution to $g_{\mu}-2$ comes from the loops
involving low mass pseudoscalar at moderate to large $\tan \beta$. When the non-diagonal elements of the Yukawa matrices are non-zero, similar diagrams will contribute to LFV decays, such as $\mu \rightarrow e \gamma$, $\tau \rightarrow e \gamma$ and $\tau \rightarrow \mu \gamma$~\footnote{$\mu \rightarrow 3 e$ and $\mu-e$ can also be important but the recent experimental bounds~\cite{Kuno:1999jp,Lindner:2016bgg}  are less stringent compared to the $\mu \rightarrow e \gamma$ process~\cite{Crivellin:2014cta}.}. Non-observation of these processes puts a strong constraint on the flavor-changing Yukawa couplings as well as the masses running in the loops and $\tan \beta$. Evidently, low mass pseudoscalar and large $\tan \beta$ are disfavored in this regard, creating a tension between these limits and the observed $g_{\mu}-2$ in our model. After careful consideration of all the low energy constraints, we identify regions of parameter space which explain the observed $g_{\mu} - 2$ and are consistent with the limits from LFV decays. However, there are various other constraints on the model parameter space and therefore it is necessary to check the validity of the region of interest in terms of all the relevant constraints. Understanding the interplay between various constraints is our main objective of this section.

\subsection{Limits from low energy measurements}

The observation of lepton flavor violation in the neutrino sector certainly motivates new physics resulting in LFV in the charged lepton sector, which can be accommodated in many BSM models. However, since no such signal has been observed yet, there are strong limits on these LFV processes~\cite{Aubert:2009ag}. Our primary interest from the $g_{\mu}-2$ requirements is the low mass $m_A$ region. Similar to muon anomaly, the LFV processes will also be dominated by the pseudoscalar contribution in the loop. Therefore these limits from the low energy LFV processes will essentially constrain the non-diagonal lepton-Yukawa couplings of the pseudoscalar $A$ (see Eq.~\ref{eq:Yu_phi}).

The strongest bound in the $\mu -e$ sector ( $ BR(\mu \to e \gamma) < 4.2 \times 10^{-13}$) comes from MEG experiment~\cite{TheMEG:2016wtm}. Similar to the $\mu-e$ sector, there are strong constraints on $(\tau \to e \gamma)$ and $(\tau \to \mu \gamma)$  branching ratio.
Current bound on $ BR(\tau \to e \gamma) < 3.3 \times 10^{-8}$~\cite{Aubert:2009ag} and  $ BR(\tau \to \mu \gamma) < 4.4  \times 10^{-8}$~\cite{Aubert:2009ag} puts a strong constraint on $y_{\tau e}$ and $y_{\mu\tau}$ respectively.

We calculate the LFV processes in one-loop as well as in two-loop~\cite{Lindner:2016bgg,Omura:2015xcg}. The flavor violating coupling between scalars and leptons at the tree-level occurs due to the presence of flavor non-diagonal Yukawa matrices in the generalized 2HDM, which in turn enables the LFV decays at one as well as two-loop. We have found that for $\tau \rightarrow \mu \gamma$ and $\tau \rightarrow e \gamma$ process, the two-loop contribution to the decay amplitudes adds up to a mere $\sim 2\%$ of their one-loop counterpart. On the contrary, in the case of $\mu \rightarrow e \gamma$, the addition of two-loop contribution induces 3 times enhancement to the one-loop amplitude.

\begin{figure}[!h]
	\centering
	\includegraphics[width=7cm,height=6cm]{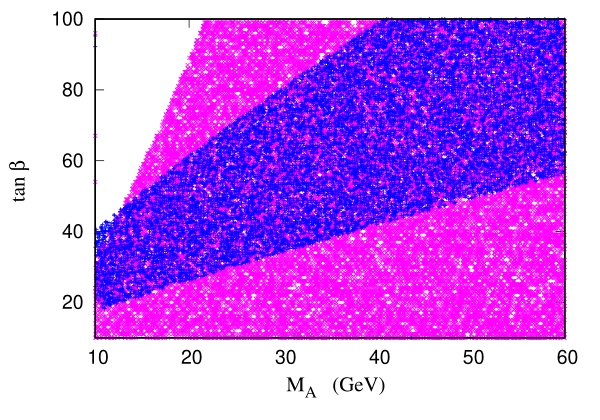}
	\includegraphics[width=7cm,height=6cm]{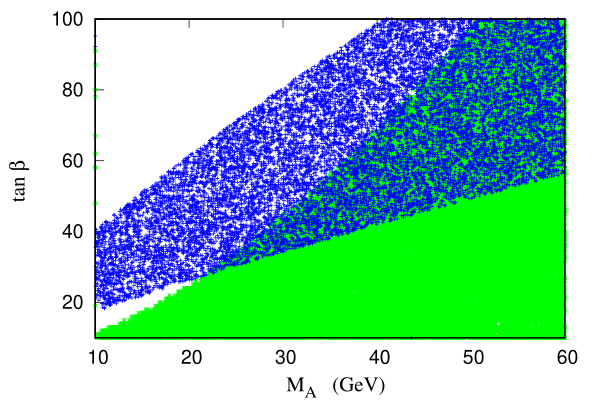}
	\includegraphics[width=7cm,height=6cm]{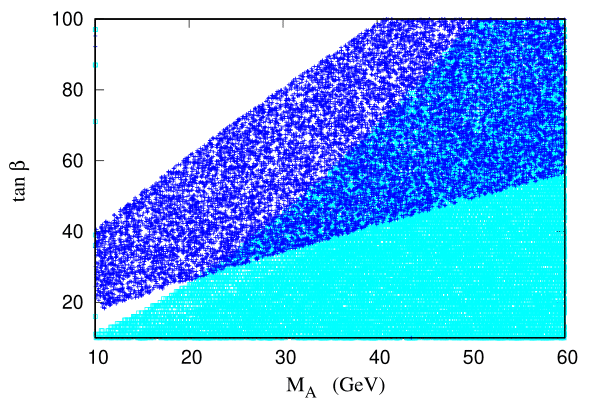}
	\caption{\it The magenta, green and cyan regions are the allowed range for $\mu \to e \gamma$,
	$\tau \to e \gamma$ and $\tau \to \mu \gamma$ respectively. The blue band is the allowed 
	3$\sigma$ allowed range for muon anomaly. The overlapping regions satisfy both constraints.
	The flavor changing couplings are taken to be $y_{\mu e}= 10^{-7},
	y_{\tau e}=4\times 10^{-5},y_{\mu \tau}=4\times 10^{-5}$. The non-standard neutral CP-even Higgs mass is 450 GeV and charged Higgs mass is 460 GeV.}
	
	\label{lfv}
\end{figure}

We have seen that BR($\tau \rightarrow e\gamma$) constrains $y_{\tau e} < 10^{-4}$ and BR($\tau \rightarrow \mu\gamma$ ) constrains
 $y_{\tau \mu} < 10^{-4}$ . However, for $y_{\mu e}$ the situation is not so straightforward. Unlike $\tau \rightarrow e \gamma$ and $\tau \rightarrow \mu \gamma$, the decay $\mu \rightarrow e \gamma$ does not primarily constrain $y_{\mu e}$ coupling as discussed earlier in detail in~\cite{Ghosh:2020tfq}. However, calculating the amplitudes at two-loop, to satisfy all the three LFV conditions simultaneously along with muon anomaly, the coupling $y_{\mu e}$ gets a strong upper bound ($ < 10^{-6}$). In Fig.\ref{lfv} we have plotted the regions allowed by LFV constraints in $m_A - \tan \beta$ plane for specific choices of flavor changing Yukawa couplings where we have also superimposed the region allowed by the recent $g_{\mu}-2$ data on the region allowed by low energy LFV data. These particular choices of flavor violating Yukawa couplings produce an adequate event rate at the HL-LHC which we will encounter shortly in section~\ref{collider}.

\subsection{Theoretical constraints}
\label{theory_constraints}

Theoretical constraints comprise perturbativity, unitarity, and vacuum stability conditions, imposed on the model parameters at the electroweak scale. Effects of these constraints on various 2HDMs have been studied in detail in the literature~\cite{Crivellin:2013wna,Arbey:2017gmh,Hussain:2017tdf}. It has been pointed out that large separation between $m_A$ and $m_{H^{\pm}}$ is disfavored from the requirement of vacuum stability and perturbativity. However, the allowed range of mass differences depends on the ``right-sign'' and ``wrong-sign'' region of 2HDM as we will see shortly.
As we have seen low $m_A$ and large $\tan \beta$ region is favored from the requirement of $g_{\mu}-2$, it is imperative to look at the allowed upper limit on $m_H^{\pm}$ for this region of parameter space. In the following, we discuss the theoretical constraints one by one, with our focus on the ``wrong-sign'' region of the parameter space.

\medskip
\noindent
$\bullet$ {\bf perturbativity and unitarity:} The requirement that 2HDM is a perturbative quantum field theory at the electroweak scale, implies all quartic couplings $C_{H_iH_jH_kH_l} < 4\pi$. Moreover, unitarity bound on the tree level scattering
amplitudes puts an upper bound on the eigenvalues of the scattering matrices $|a_i|\leq 16\pi$ .

For our upcoming discussion, it will be useful to express the physical masses of the scalars in terms of the quartic couplings in the following manner~\cite{Gunion:2002zf}. 

\begin{eqnarray}
m_A^2 &=& \frac{m_{12}^2}{\sin\beta \cos\beta} - \frac{1}{2}(2\lambda_5 + \frac{\lambda_6}{\tan\beta} + \lambda_7 \tan\beta) v^2  \\
m_{H^{\pm}}^2 &=& m_A^2 + \frac{1}{2} v^2 (\lambda_5 - \lambda_4)
\label{massdiff}
\end{eqnarray}

It is clear from Eq.~\ref{massdiff} that $m_{H^{\pm}}^2 - m_A^2$ is proportional to $\lambda_5 - \lambda_4$ which should be less than $ \lambda_3 + \sqrt{\lambda_1 \lambda_2}$ from the requirement of vacuum stability (see Eq.~\ref{stability}). Therefore these conditions along with the requirement of perturbativity ie. $C_{H_iH_jH_kH_l} < 4\pi$ puts an upper limit on the mass square difference  $m_{H^{\pm}}^2 - m_A^2 < 4\pi v^2$, which implies $m_H^{\pm} \lsim 870$ GeV for very low $m_A$.

\medskip

We will now proceed further to discuss the effect of the theoretical constraints applied on the ``wrong-sign'' region of the parameter space. We can write the quartic couplings in terms of physical mass parameters, $m_{12}^2$ and hard $Z_2$-breaking parameters $\lambda_6$ and $\lambda_7$~\cite{Gunion:2002zf}.

\begin{eqnarray}
\nonumber && \lambda_1 = \frac{m_H^2c_\alpha^2+m_h^2s_\alpha^2-m_{12}^2t_{\beta}}{v^2c_\beta^2} - \frac{3}{2}\lambda_6 t_{\beta} + \frac{1}{2}\lambda_7 t_{\beta}^2,\\
\nonumber && \lambda_2 = \frac{m_H^2s_\alpha^2+m_h^2c_\alpha^2-m_{12}^2t_\beta^{-1}}{v^2s_\beta^2} + \frac{1}{2}\lambda_6 t_{\beta}^{-3} - \frac{3}{2}\lambda_7 t_{\beta}^{-1},\\
\nonumber && \lambda_3 = \frac{(m_H^2-m_h^2)c_\alpha s_\alpha+2m_{H^\pm}^2s_\beta c_\beta-m_{12}^2}{v^2s_\beta c_\beta} - \frac{1}{2}\lambda_6 t_{\beta}^{-1} - \frac{1}{2}\lambda_7 t_{\beta},\\
\nonumber && \lambda_4 = \frac{(m_A^2-2m_{H^\pm}^2)s_\beta c_\beta+m_{12}^2}{v^2s_\beta c_\beta} - \frac{1}{2}\lambda_6 t_{\beta}^{-1} - \frac{1}{2}\lambda_7 t_{\beta},\\
&& \lambda_5 = \frac{m_{12}^2 - m_A^2s_\beta c_\beta}{v^2s_\beta c_\beta} - \frac{1}{2}\lambda_6 t_{\beta}^{-1} - \frac{1}{2}\lambda_7 t_{\beta}.
\label{eq:paratran}
\end{eqnarray}

It is clear from the expression of $\lambda_1$ in Eq.~\ref{eq:paratran} that when $m_H >> m_h$, to have $\lambda_1$ in the perturbative limit, the soft $Z_2$ breaking parameter $m_{12}^2 \approx \frac{m_H^2}{\tan \beta}$. We note here that, when $\lambda_6, \lambda_7$ are non-zero, larger deviation from this limit is allowed, as compared to the case, $\lambda_6, \lambda_7 \approx 0$.

\medskip
\noindent
$\bullet$ {\bf Vacuum stability:} The vacuum stability demands there can exist no direction in the field space in which $\cal V \rightarrow -\infty$. This implies the following conditions on the quartic couplings of the Higgs potential~\cite{Ferreira:2004yd,Ferreira:2009jb}\footnote{The work of ~\cite{Ivanov:2006yq} provides all the necessary and sufficient conditions for vacuum stability in the presensence of non-zero $\lambda_6$ and $\lambda_7$. However, the relations that can be derived in that case are complicated and not very illuminating~\cite{Ferreira:2009jb}. Therefore, here we have mentioned only the necessary (not sufficient) condition in Eq.~\ref{stability}.}.

\begin{eqnarray}
\label{vs1}
\lambda_{1,2} > 0 \,, \\
\label{vs2}
\lambda_3  > -\sqrt{\lambda_1 \lambda_2} \,\\
\label{vs3}
|\lambda_5| < \lambda_3 + \lambda_4 + \sqrt{\lambda_1 \lambda_2}\,\\
\label{vs4}
2|\lambda_6 + \lambda_7| < \frac{\lambda_1 + \lambda_2}{2} + \lambda_3 + \lambda_4 + \lambda_5 
\label{stability}
\end{eqnarray} 

The condition in Eq.~\ref{vs3} can be rewritten as $\lambda_{3}+\lambda_4-\lambda_5>-\sqrt{\lambda_1\lambda_2} $ for $m_H>m_A$.

One of the key features of this model is that the upper limit on the heavy Higgs mass show quite different behavior in the ``wrong-sign'' region as compared to the ``right-sign'' limit of the  Yukawa coupling $y_{h}^{f_i}$ \cite{Ferreira:2014naa}. It is obvious that since we are interested in the upper limit on the heavier CP-even neutral scalar, it will suffice to discuss Scenario 1, ie. $m_h = 125$ GeV. In this case the ``wrong-sign'' region implies $y_{h}^{\ell} \times \sin(\beta - \alpha) \approx -1$ as we have seen in the previous section. 
Using Eq.~\ref{yukawa1} and \ref{eq:paratran} one can derive the following relation~\cite{Chun:2015hsa}.

\begin{eqnarray}
\lambda_{3}+\lambda_4-\lambda_5 = 
{ 2 m_A^2 + (y_{h}^{\ell} -\frac{\cos(\beta-\alpha) \Sigma^\ell}{\sqrt{2} \cos\beta}) \sin({\beta-\alpha}) m_h^2 - (\sin^2(\beta-\alpha) + (y_{h}^{\ell} - \frac{\cos(\beta-\alpha) \Sigma^\ell}{\sqrt{2} \cos\beta}) s_{\beta-\alpha}) m_H^2 \over v^2}  \nonumber \\
 - \frac{1}{2}\lambda_6 t_{\beta}^{-1} - \frac{1}{2}\lambda_7 t_{\beta} +{\cal O}({1\over t^2_\beta}) 
 \label{lam345}
\end{eqnarray}

\begin{figure}[!hptb]

\centering
\includegraphics[width=7.5cm, height=6.5cm]{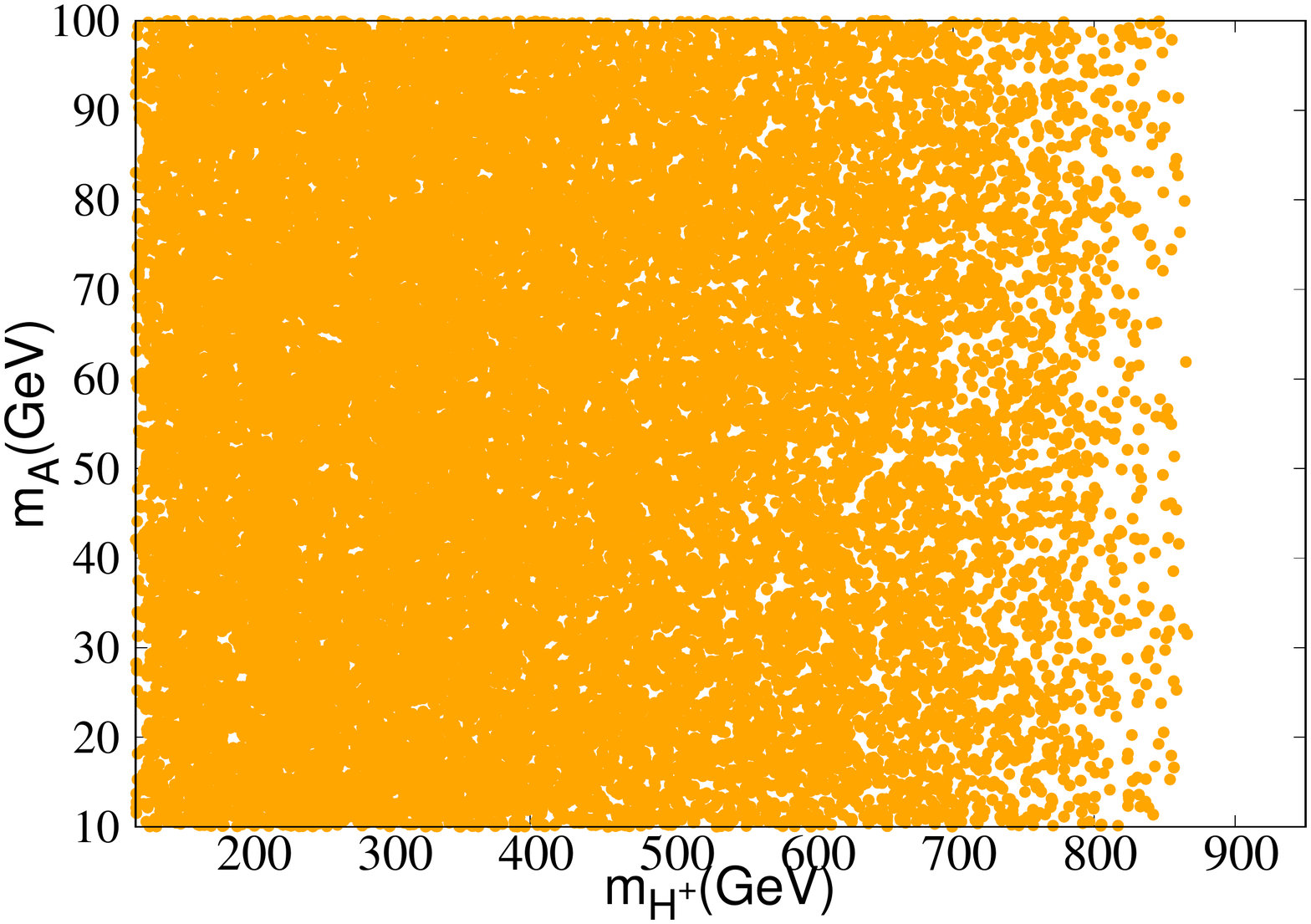}
\caption{{\it Allowed parameter space in $m_{H^{\pm}}-m_A$ plane consistent with theoretical bounds.}}
\label{pert_bound_ws}
\end{figure}

We can see that when $y_{h}^{\ell} s_{\beta-\alpha} \approx +1$ which is the ``right-sign'' alignment case, this condition sets a strong upper bound on $m_H$~\cite{Broggio:2014mna}. However, one can see from Eq.~\ref{lam345} and ~\ref{vs3}, that if in the large $\tan \beta$ limit, $\lambda_7$ is taken to be non-zero positive values, the upper limit on $m_H$, from stability criteria, becomes stronger, whereas for negative $\lambda_7$, it becomes weaker. 
On the other hand, in the ``wrong-sign'' limit $y_{h}^{\ell} s_{\beta-\alpha} = -1$. Here the coefficients of $m_H^2$ term cancel naturally and arbitrarily large $m_H$ is allowed by the stability criteria. The terms involving $\lambda_6$ and $\lambda_7$ only contribute a small quantity($\sim{\cal O}(1))$ as long as their values are small.  The region allowed by stability and perturbativity criteria has been shown in Fig.~\ref{pert_bound_ws}. 
For this purpose, we have performed a scan in the following range of parameters for scenario 1, where $m_h = 125$ GeV and hard $Z_2$-symmetry breaking parameters $\lambda_6$ and $\lambda_7$ are assumed to be non-zero.  

 $m_A \in$ [10.0 GeV, 60.0 GeV],~~$m_H \in$ [126 GeV, 1 TeV],~~$m_H^{\pm} \in$ [89.0 GeV, 1 TeV],~~$m_{12}^2 \in$ [$-10^5$ GeV$^2$, $10^5$ GeV$^2$]$,~~\tan \beta \in$ [10, 70],~~$|\sin(\beta - \alpha)| \in$ [0.99, 1],~~$\lambda_6 \in$ [0, 0.1],~~$\lambda_7 \in$ [0, 0.1]

We would like to mention here that, we have used the 2HDMC-1.8.0~\cite{Eriksson:2009ws} package to check the condition for perturbativity, unitarity, and vacuum stability for the scanned points.

\subsection{Electroweak constraints}

\begin{figure}[!hptb]
	\centering
	\includegraphics[width=8cm,height=7cm]{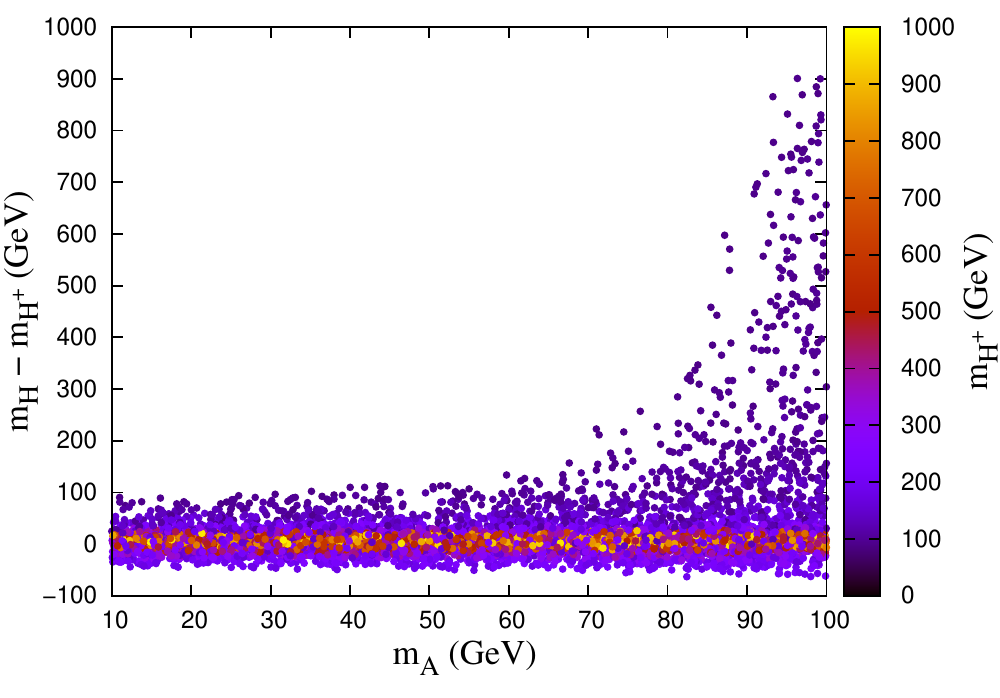}
	\caption{\it Parameter space satisfying electroweak constraints in the plane of $m_A$ and $m_H - m_{H^{\pm}}$ with $m_{H^{\pm}}$ as the color-axis.}
	\label{stu_3d}
\end{figure}

The custodial SU(2) is a symmetry of the SM Higgs potential and can be broken at the loop level in 2HDM. Electroweak precision measurements of the oblique parameters, namely $S, T, U$ parameters have been conducted by the Gfitter group~\cite{BAAK:2014gga}. The experimental values of electroweak observables within experimental error can restrict $|\Delta m|= |m_H - m_{H^\pm}|$ depending on $m_A$ and values of $m_{H^\pm}$~\cite{Broggio:2014mna}. The status of two Higgs doublet models in the light of global electroweak data has been presented in~\cite{Haller:2018nnx}. We present here the resulting allowed region in $m_A - \Delta m$ plane where $m_H^{\pm}$ has been represented as the third(color)-axis in Fig.~\ref{stu_3d}. We mention here that we have considered the elliptic contour computed with $U$ as a free parameter. This choice leaves us with a less constrained parameter space than the scenario when $U$ is fixed at 0.

We have shown in Fig.~\ref{stu_3d} the pseudoscalar mass range of our interest ($m_A \lsim 100$ GeV). It is clear from the figure that for $m_H < m_{H^{\pm}}$, it is possible to attain up to $|\Delta m| \sim 50$ GeV, in the limit $m_H^{\pm} \lsim 200$ GeV. When $m_H > m_{H^{\pm}}$, ie. $\Delta m > 0$, it is possible to get a mass gap as large as 1 TeV when $m_A$ and $m_{H^{\pm}}$ are almost degenerate. This behavior can be clearly confirmed from the calculation of $S$ and $T$ parameter~\cite{He:2001tp,Grimus:2007if,Bhattacharyya:2015nca}.

\subsection{Constraints from $h_{SM} \rightarrow AA$ search at the LHC}

As our study focuses on low mass pseudoscalar, the most crucial collider constraint comes from the direct search for SM-like Higgs boson decaying into a pair of pseudoscalars. One should note that BR($h_{SM} \rightarrow AA$) depends on the scenario (whether $m_h$ or $m_H$ is 125 GeV) and also on the ``right-sign'' or ``wrong-sign'' region.

First, we consider Scenario 1 ie. $m_h = 125$ GeV. The partial decay width of Higgs decaying to a pair of pseudoscalars is given by
\begin{equation}
\Gamma(h\rightarrow AA) = \frac{1}{32\pi}\frac{g_{hAA}^2}{m_h}\sqrt{1-4m_A^2/m_h^2}
\label{widthhaa}
\end{equation}

Using the relations between the quartic couplings $\lambda'$s and the physical masses and Higgs mixing parameter $m_{12}^2$, in the alignment limit $|\sin(\beta - \alpha)| \approx 1$ and with large $\tan \beta$~\footnote{The wrong-sign region does not imply alignment, and exact alignment $|\sin(\beta - \alpha)|=1$ can never achieve wrong-sign criteria. However, if one goes to large $\tan \beta (\gsim 15)$, the requirement that $|y_{h}^{\ell}|\approx 1$, forces us to choose $|\sin(\beta - \alpha)|>0.99$.}, one can find the $hAA$ coupling~\cite{Gunion:2002zf} as the following.

\begin{equation}
g_{hAA} \propto (\lambda_3 + \lambda_4 - \lambda_5 )v \approx  \frac{\sin(\beta - \alpha){y_h^{\ell}}({m^2_h} - {m^2_H}) + 2{m^2_A} - m_{12}^2/s_{\beta}c_{\beta}}{v} - \frac{v}{2}(\lambda_6 t_{\beta}^{-1} + \lambda_7 t_{\beta})
\end{equation}
Expressing the quantity $y_h^{\ell}\sin(\beta - \alpha)$ in terms of $g_{hAA}$, $\lambda_6, \lambda_7$ and mass parameters, we get
\begin{equation}
{y_h^{\ell}}\sin(\beta - \alpha) = \frac{g_{hAA}v + m_{12}^2/s_{\beta}c_{\beta} + (\lambda_6 t_{\beta}^{-1} + \lambda_7 t_{\beta})v^2/2 - 2m_A^2}{m_h^2 - m_H^2}
\label{ylhwrong}
\end{equation}
We can see from Eq.~\ref{widthhaa} that when $m_A \lsim \frac{m_h}{2}$, the only way a small branching ratio for BR($h \rightarrow AA)$ can be achieved is when the coupling $g_{hAA}$ is extremely small. 
We should also remember from our discussion of perturbativity that in this scenario $m_{12}^2 \approx \frac{m_H^2}{\tan \beta}$ to ensure perturbativity of quartic couplings. 
Therefore, if we demand perturbativity and impose the condition $g_{hAA} \approx 0$, Eq.~\ref{ylhwrong} implies ${y_h^{\ell}}\sin(\beta - \alpha) < 0$, as long as $\lambda_6, \lambda_7$ are taken to be small (which is the case in our range of scan). On the other hand, ${y_h^{\ell}}\sin(\beta - \alpha) > 0$ will lead to large negative $g_{hAA}$, which is not desirable. In other words, ``wrong-sign'' lepton-Yukawa coupling is more favored in Scenario 1 in order to satisfy the small $h \rightarrow AA$ branching ratio as well as perturbativity of quartic couplings in the chosen range of our scan. It is worth mentioning that, if $\lambda_6$ and $\lambda_7$(most importantly $\lambda_7$ in the large $\tan \beta$ region) are chosen to be negative, it is possible to achieve right-sign region which will yield $g_{hAA} \approx 0$ and will respect perturbativity. In that case, one will require large $|\lambda_7|$ as $m_H$ increases, to get $|{y_h^{\ell}}\sin(\beta - \alpha)| \approx 1$. However, the phenomenology of the low mass pseudoscalar that we are interested in will not be affected by this choice and therefore we have not explicitly explored this region in this work.

The other possibility is to consider the case when the heavier CP even Higgs is SM-like, ie $m_H = 125$ GeV which is our Scenario 2.
Here the decay width of 125-GeV Higgs decaying to a pair of pseudoscalars is given by

\begin{equation}
\Gamma(H\rightarrow AA) = \frac{1}{32\pi}\frac{g_{HAA}^2}{m_h}\sqrt{1-4m_A^2/m_H^2}
\label{widthHaa}
\end{equation}

Here too, like the previous scenario, the limit on BR($H \rightarrow AA$) will indicate an extremely small value of the coupling $g_{HAA}$, whose expression in the alignment limit ie. $|\cos(\beta -\alpha)| \approx 1$ is given as follows: 

\begin{equation}
g_{HAA} \propto (\lambda_3 + \lambda_4 - \lambda_5 )v \approx  \frac{\cos(\beta - \alpha){y_H^{\ell}}({m^2_H} - {m^2_h}) + 2{m^2_A} - m_{12}^2/s_{\beta}c_{\beta}}{v} - \frac{v}{2}(\lambda_6 t_{\beta}^{-1} + \lambda_7 t_{\beta})
\end{equation}

Expressing the quantity $y_H^{\ell}\cos(\beta - \alpha)$ in terms of $g_{HAA}$, $\lambda_6, \lambda_7$ and mass parameters we get
\begin{equation}
{y_H^{\ell}}\cos(\beta - \alpha) = \frac{g_{HAA}v + m_{12}^2/s_{\beta}c_{\beta}  + (\lambda_6 t_{\beta}^{-1} + \lambda_7 t_{\beta})v^2/2 - 2m_A^2}{m_H^2 - m_h^2}
\label{ylHrght}
\end{equation}

We can see that, as we are concerned with low pseudoscalar mass here($m_A \lsim \frac{m_H}{2}$), in the limit $g_{HAA} \approx 0$, $m_{12}^2 \approx \frac{m_H^2}{\tan \beta}$ and small $\lambda_6, \lambda_7$, one is naturally bound to choose $y_H^{\ell}\cos(\beta - \alpha) > 0$ ie. ``right-sign'' in Scenario 2~\cite{Ghosh:2020tfq}. In this work, we will focus on the ``wrong-sign'' sector and therefore we will consider only Scenario 1 ie. $m_h = 125$ GeV. In Scenario 2, wrong-sign can in principle be achieved with sufficiently large negative values of $\lambda_6$ or $\lambda_7$, a possibility we are not considering in this work.

\subsection{B-physics constraints}

From our model description detailed in Section~\ref{model}, we have seen that the charged Higgs couplings to quarks and leptons are modified in the presence of flavor-changing terms in the Yukawa Lagrangian. That leads to rare processes involving $B-$mesons. However, the free parameters of the model receive strong constraints by the experimental bounds on the rare FCNC processes. While the FCNC within the first two generations is naturally suppressed by the small quark masses, substantial freedom is still allowed in the third generation quark sector~\cite{Ghosh:2020tfq}. Therefore, we have taken only $\lambda_{tt}$ and $\lambda_{bb}$ to be non-zero, where $\lambda_{tt}$ and $\lambda_{bb}$ are defined as the $H t \bar t$ and $H b \bar b$ coupling strengths respectively.

The strongest and most relevant limit in this context comes from the $B \rightarrow X_s \gamma$ decay. The impact of these constraints on the parameter space of various 2HDMs has been studied in great detail in earlier works~\cite{Crivellin:2013wna,Arbey:2017gmh,Hussain:2017tdf,Crivellin:2019dun}. In two Higgs doublet models, a crucial additional contribution to $B \rightarrow X_s \gamma$ comes from the charged Higgs boson-top quark penguin diagrams and its contribution depends on $m_{H^{\pm}}$. In the type X 2HDM, the charged Higgs penguin diagram's contribution interferes destructively with its SM counterpart and gives negligible additional contribution at large $\tan \beta$. Therefore, in Type X case, no strong constraint appears on the mass of the charged Higgs boson. As our model can be perceived as a perturbation about the type X scenario, even in the presence of non-zero FCNC Yukawa matrix elements, one can get low enough $m_{H^{\pm}}$~\cite{Xiao:2003ya,Mahmoudi:2009zx,Arhrib:2017yby,Cherchiglia:2017uwv,Enomoto:2015wbn} with suitably chosen $\lambda_{tt}$ and $\lambda_{bb}$ couplings. In our analysis $\lambda_{tt} \sim 0.5$ and $\lambda_{bb} \sim 12$, which allows a charged Higgs mass $m_{H^{\pm}} \gtrsim 250$ GeV. For our analysis, we have kept $m_{H^{\pm}} = 460$ GeV. The non-standard CP-even scalar mass($m_H$) is chosen to be 450 GeV obeying the allowed mass gap (see Fig.~\ref{stu_3d}). We would like to mention here that, for low $m_A$, such large charged Higgs mass is allowed only in case of ``wrong-sign'' Yukawa coupling as discussed in the Section~\ref{theory_constraints}. Therefore compared to the ``right-sign'' region~\cite{Ghosh:2020tfq}, one can choose larger $\lambda_{bb}$ coupling which will enhance the production cross-section of the pseudoscalar Higgs boson and is of paramount interest in the collider study of this scenario. We will see the effect of this choice of parameters in our discussion of collider analysis in the next section.

\section{Collider Searches}\label{collider}

From the discussions of the preceding sections, it is clear that flavor violation in the lepton-Yukawa sector will result in flavor-violating decays of $\mu$ and $\tau$ leptons. These decays are induced at loop level by the tree-level flavor-violating couplings between the scalars and the leptons. These tree-level flavor-violating Yukawa couplings can be probed at the collider experiments~\cite{Banerjee:2016foh,Primulando:2016eod,Primulando:2019ydt,Jana:2020pxx}.

We explore the decay of the CP-odd scalar $A$ in flavor violating leptonic modes at the HL-LHC, in the context of generalized 2HDM with ``wrong-sign'' lepton-Yukawa coupling, motivated by its unique phenomenology. The relevant signal process is the following.

\begin{equation}
 p p \to A \to \ell \tau_{\ell'}
\end{equation}

\noindent
Where $\ell, \ell'= e,\mu$ and $\tau_{\ell'}$ stand for the leptonic decay of $\tau$. Therefore the final state of our interest is $\ell^{+}\ell'^{-} + \slashed{E_T}$. 

The SM backgrounds that give rise to similar final states consist of $\tau\tau/ e e / \mu\mu ,  t \bar{t}, W^{\pm}$+jets,
di-boson, SM Higgs~\cite{Banerjee:2016foh, Sirunyan:2019shc}. The major and irreducible background turns out to be the leptonic final state of $\tau \tau$.  
 $t\bar{t}$(leptonic) also contributes substantially due to its large production cross-section. $t \bar t$ semileptonic and $W+$ jets background, despite having significant cross-section, end up with reduced contribution after application of our preselection cuts. Therefore for our purpose it will suffice to consider the leptonic mode of $\tau \tau$ and $t \bar t$ backgrounds. The $ee/\mu\mu$ background poses a threat due to the enormous production cross-section. However, we have checked that in our signal region, this background contributes $\lsim$ 5\% of the $\tau \tau$ background and therefore plays a sub-dominant role. The di-boson and SM Higgs background turn out to be insignificant compared to the aforementioned processes due to much smaller production cross-section~\footnote{For validation, we have generated the backgrounds for $\sqrt{s}$ = 13 TeV using Madgraph@NLO~\cite{Alwall:2014hca} and
compared with the background event expectation as given in the CMS paper~\cite{Sirunyan:2019shc}. We have found that the backgrounds are consistent with the CMS background numbers up to 97\% .}.

We chose a few benchmark points obeying all the experimental and theoretical constraints. As the branching ratios of the pseudoscalar decaying to flavor violating final states are strongly constrained (BR($A \to \mu \tau) \approx$ BR($A \rightarrow \tau e) \approx 10^{-7}$) by the low energy LFV data, in order to have any detectable signal at the colliders we have to consider low mass pseudoscalar, which will have substantial production cross-section. We highlight the fact that for the same pseudoscalar mass it is possible to achieve a larger production cross-section in the ``wrong-sign'' region compared to the ``right-sign'' case~\cite{Ghosh:2020tfq}. The reason behind this is that the $\lambda_{bb}$ coupling which plays a crucial role in the production of the pseudoscalar can take larger value allowed by the B-physics constraints in the ``wrong-sign'' region, as in this case, one can have charged Higgs mass on the higher side.

\begin{table}[!hptb]
\centering
 \begin{tabular}{|c|c|c|c|c|c|c|c|c|c|}
  \hline
  & $\tan\beta$ &  $m_{A}$ & $m_H$ & $m_H^{\pm}$ & $m_{12}^2$ & $\lambda _6$ & $\lambda_7$ & $|\sin(\beta-\alpha)|$ &$\sigma_{prod}(\sqrt{s}=14 ~\rm{TeV})$\\
  &  &  (in GeV) & (in GeV) & (in GeV) & (in GeV$^2$) & & & & (in fb) \\
  \hline
  BP1 & 18 & 21 & 450 & 460 & 11210 & 0.001 & 0.002 & 0.994 & 0.51 \\ \hline
  BP2 & 20 & 26 & 450 & 460 & 10125 & 0.01 & 0.001 & 0.995 & 0.46 \\  \hline
  BP3 & 25 & 30 & 450 & 460 & 8100 & 0.01 & 0.0009 & 0.997 & 0.28 \\ \hline 
  BP4 & 28 & 35 & 450 & 460 & 7232 & 0.0006 & 0.0007 & 0.997 & 0.16 \\ \hline 
  BP5 & 30 & 40 & 450 & 460 & 6750 & 0.01 & 0.0005 & 0.998 & 0.096 \\ \hline 
  \hline
 \end{tabular}
	\caption{\it Benchmark points allowed by all constraints and the corresponding production cross-section of
	our signal at LO at 14 TeV LHC. }
	\label{tab:bp}
 \end{table}

In the following subsection we will present the cut-based analysis. We will perform an improved analysis using Artificial Neural Network (ANN) thereafter.

\subsection{Cut-based Analysis}

The signal and background events are generated at the leading order (LO) in {\tt Madgraph5@NLO}~\cite{Alwall:2014hca} using the {\tt NNPDF3.0} parton distributions~\cite{Ball:2014uwa}. 
Parton shower and hadronization are performed using the built-in {\tt Pythia}
\cite{Sjostrand:2006za} within Madgraph. Detector simulation is taken care of by {\tt Delphes}(v3)~\cite{deFavereau:2013fsa}. For jet formation we have used the anti-$K_{T}$ jet algorithm with jet radius $\Delta R = 0.5$.  

At the generation level the following generation-level cuts are implemented:
\begin{eqnarray}
 p_T(j,b) &>& 20 ~{\rm GeV}\,; \quad |\eta(j)| < 4.7 \,; \quad |\eta(b)| < 2.5 \,, \nonumber \\ 
 p_T(\ell) &>& 10 ~{\rm  GeV}\,, \quad |\eta(\ell)| <2.5 \,.
 \label{basic_cut}
\end{eqnarray}

Along with that, we apply the following selection cuts on certain kinematical observables which we will discuss in detail in the following.

 \begin{figure}[!hptb]
 	\centering
 	\includegraphics[width=7.2cm,height=6.0cm]{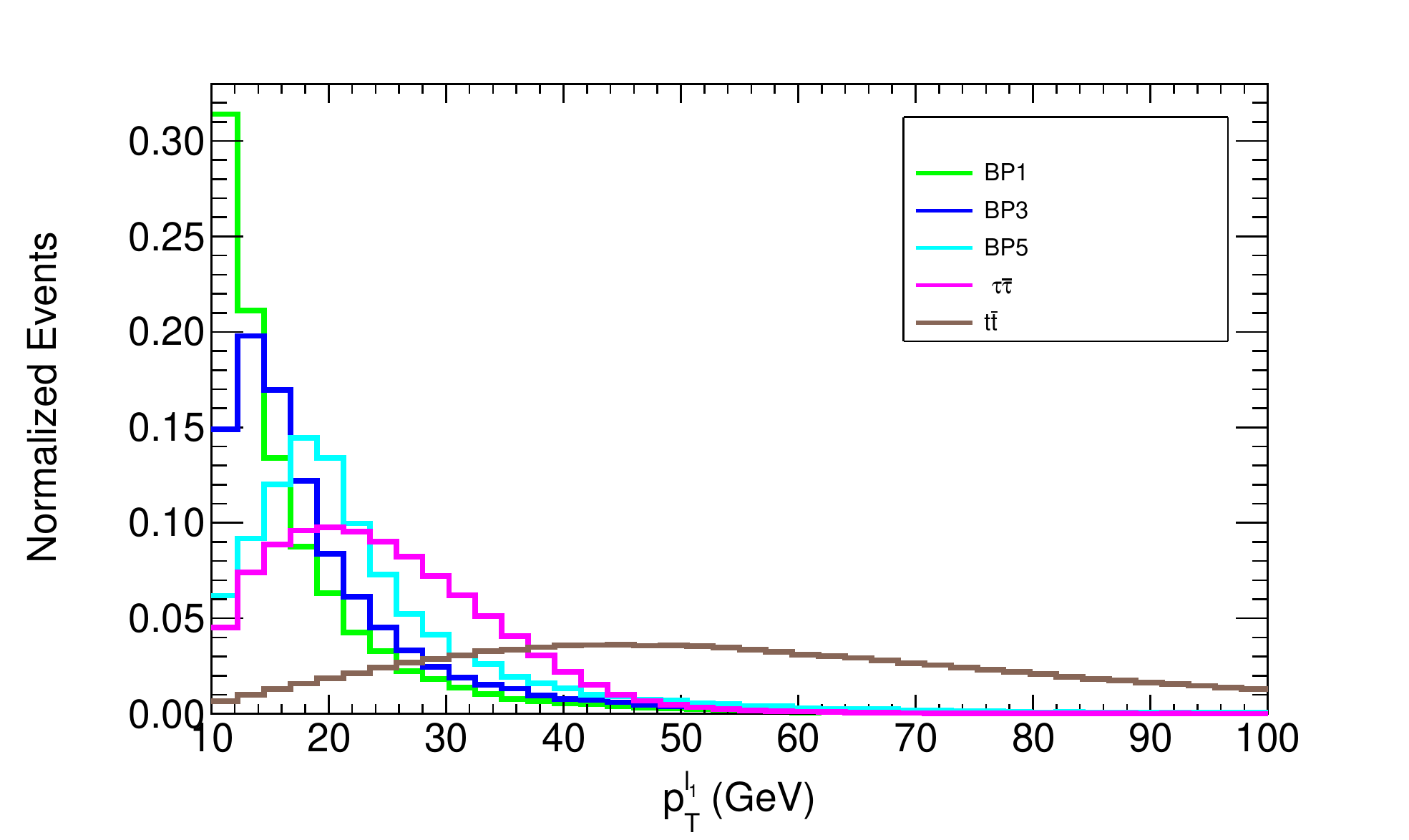}
 	\includegraphics[width=7.2cm,height=6.0cm]{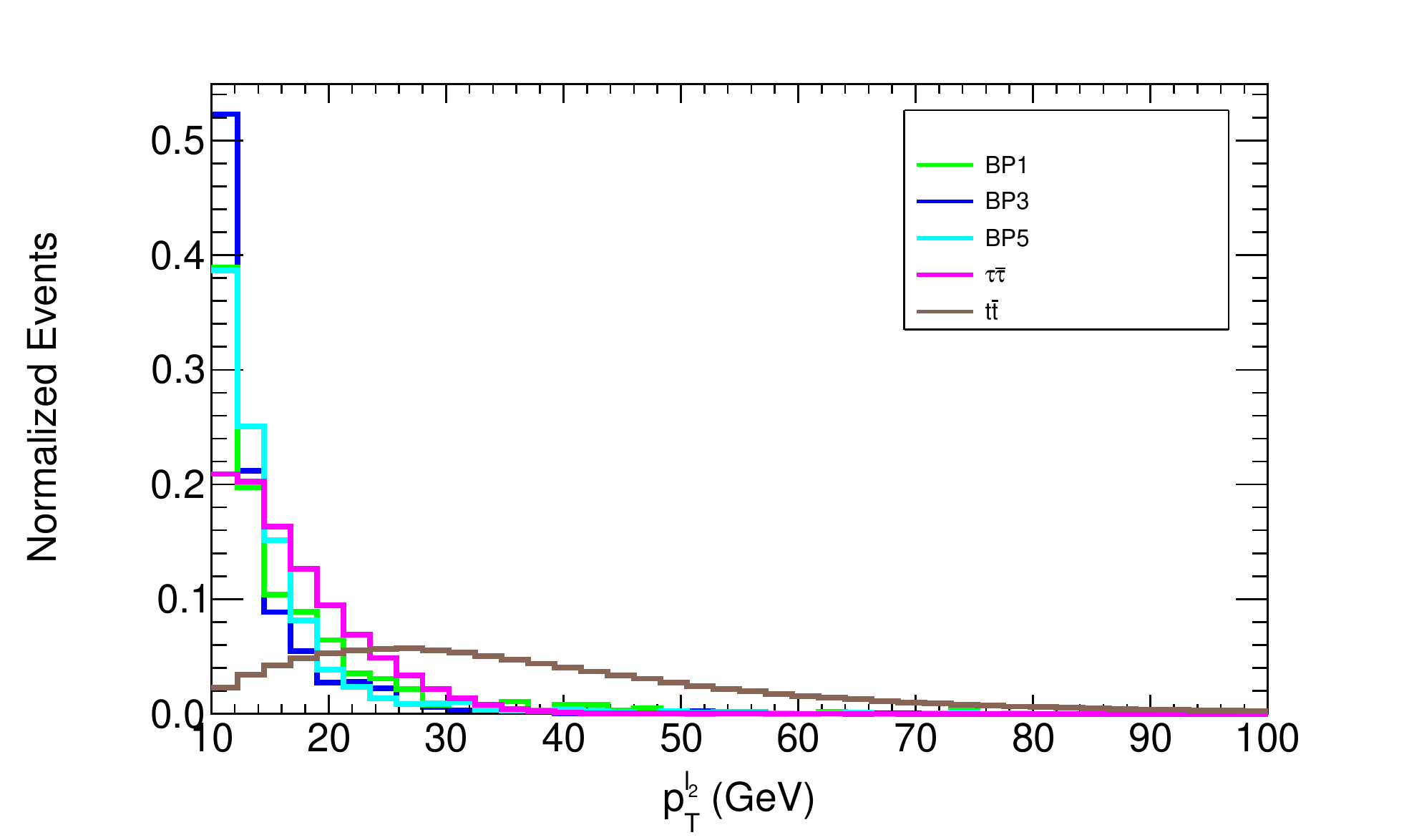}
 \caption{\it Distribution of transverse momenta of leading (left) and sub-leading (right) leptons for signal and backgrounds.}
 \label{ptlep}
 \end{figure}

\begin{itemize}
           \item {\bf $p_T$ of the leptons:} In Fig.~\ref{ptlep}, we present the $p_T$ distribution of the leading and sub-leading leptons. As the leptons in the case of signal come from the decay of low mass pseudoscalar, they show similar behavior to the leptons that are coming from the leptonic $\tau \tau$ background. Due to such overlap between signal and background, it is very difficult to put any hard $p_T$ cut on the leptons. However, we demand exactly two leptons with $p_T(\ell) >$ 10 GeV in the final state. 
Moreover, we put a $b$-veto (reject any $b-$jet with $p_T > 20$ GeV)  and jet-veto (reject any light jet with $p_T > 20$ GeV).  These particular cuts help us reduce the $t\bar{t}$ semileptonic and $W^{\pm}+$ jets background to a large extent. These are referred to as our preselection cuts in Table~\ref{tab:sig}.

 \begin{figure}[!hptb]
 	\centering
 	\includegraphics[width=7.2cm,height=6.0cm]{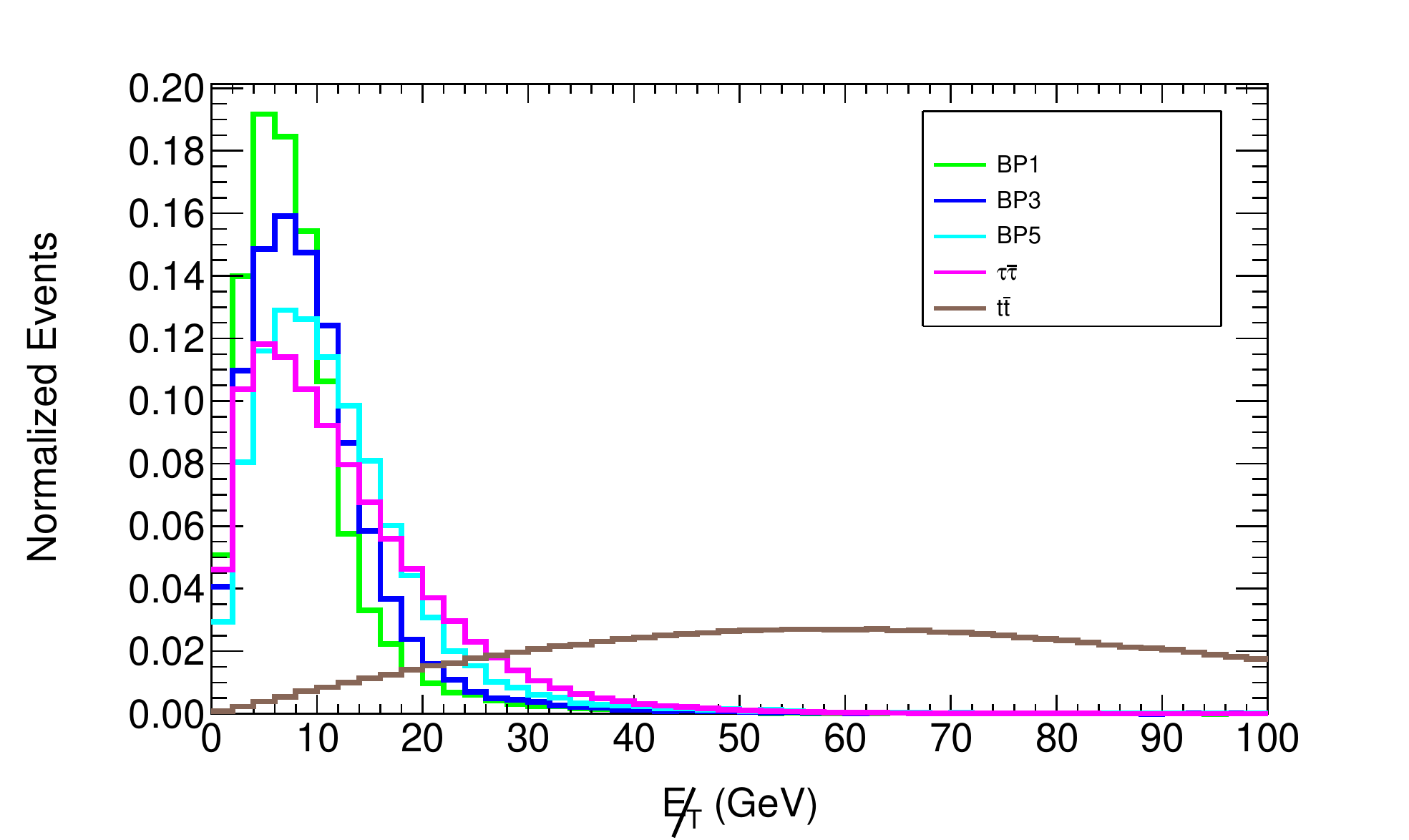}
 	\includegraphics[width=7.2cm,height=6.0cm]{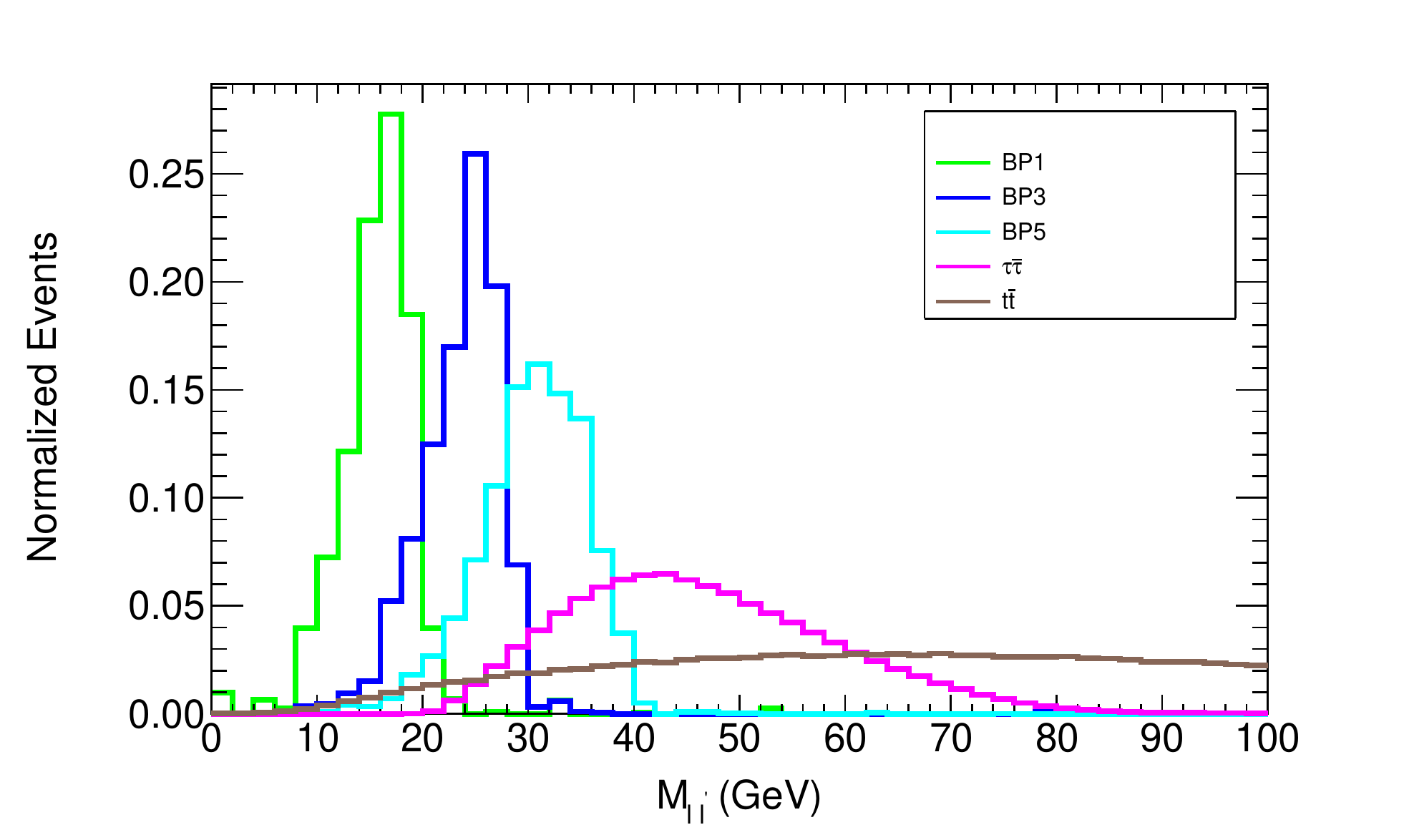}
 \caption{\it Distribution of $\slashed{E_T}$ (left) and invariant mass of two leptons for signal and backgrounds.}
 \label{met_invll}
 \end{figure}

 \begin{figure}[!hptb]
 	\centering
 	\includegraphics[width=7.2cm,height=6.0cm]{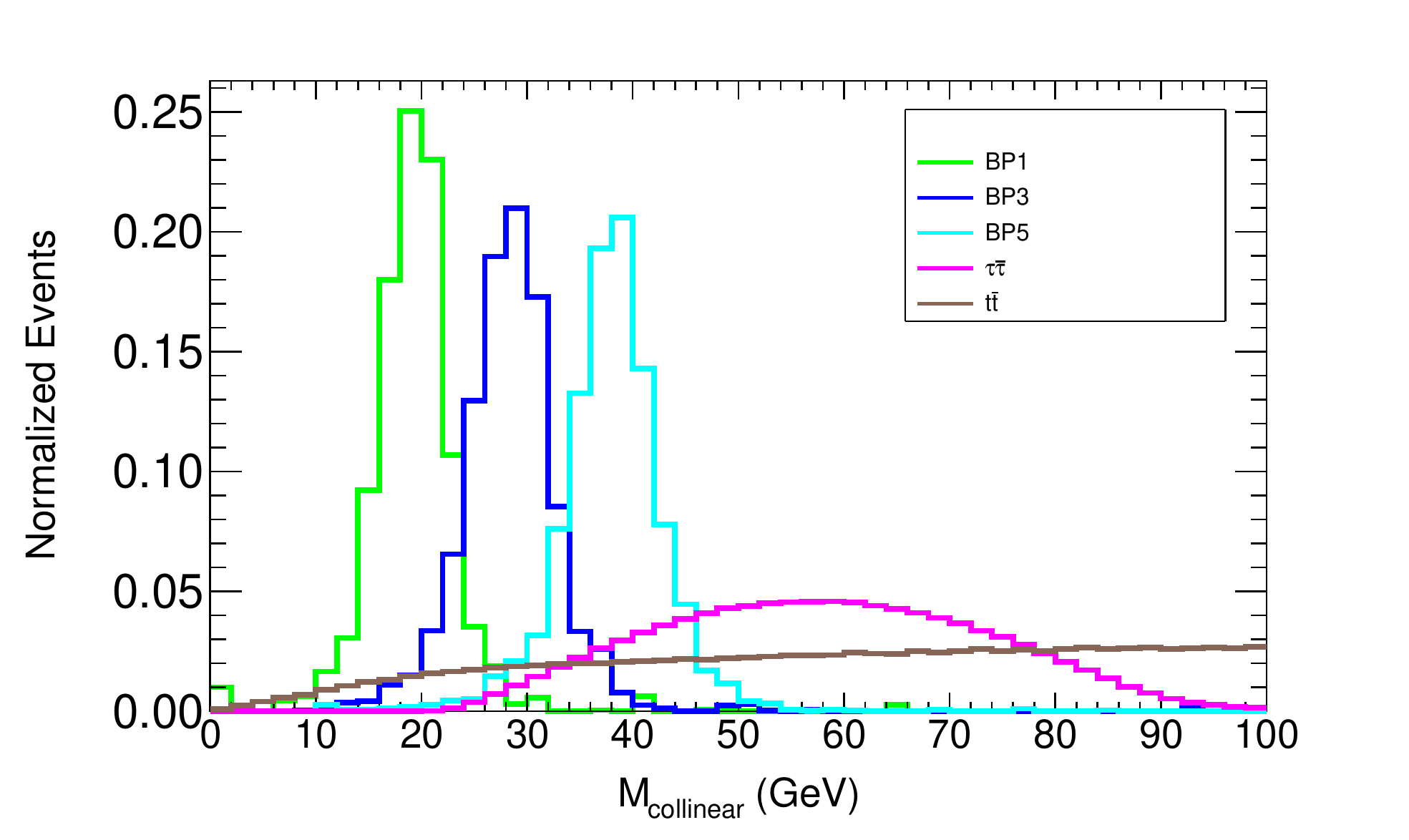}
 	\includegraphics[width=7.2cm,height=6.0cm]{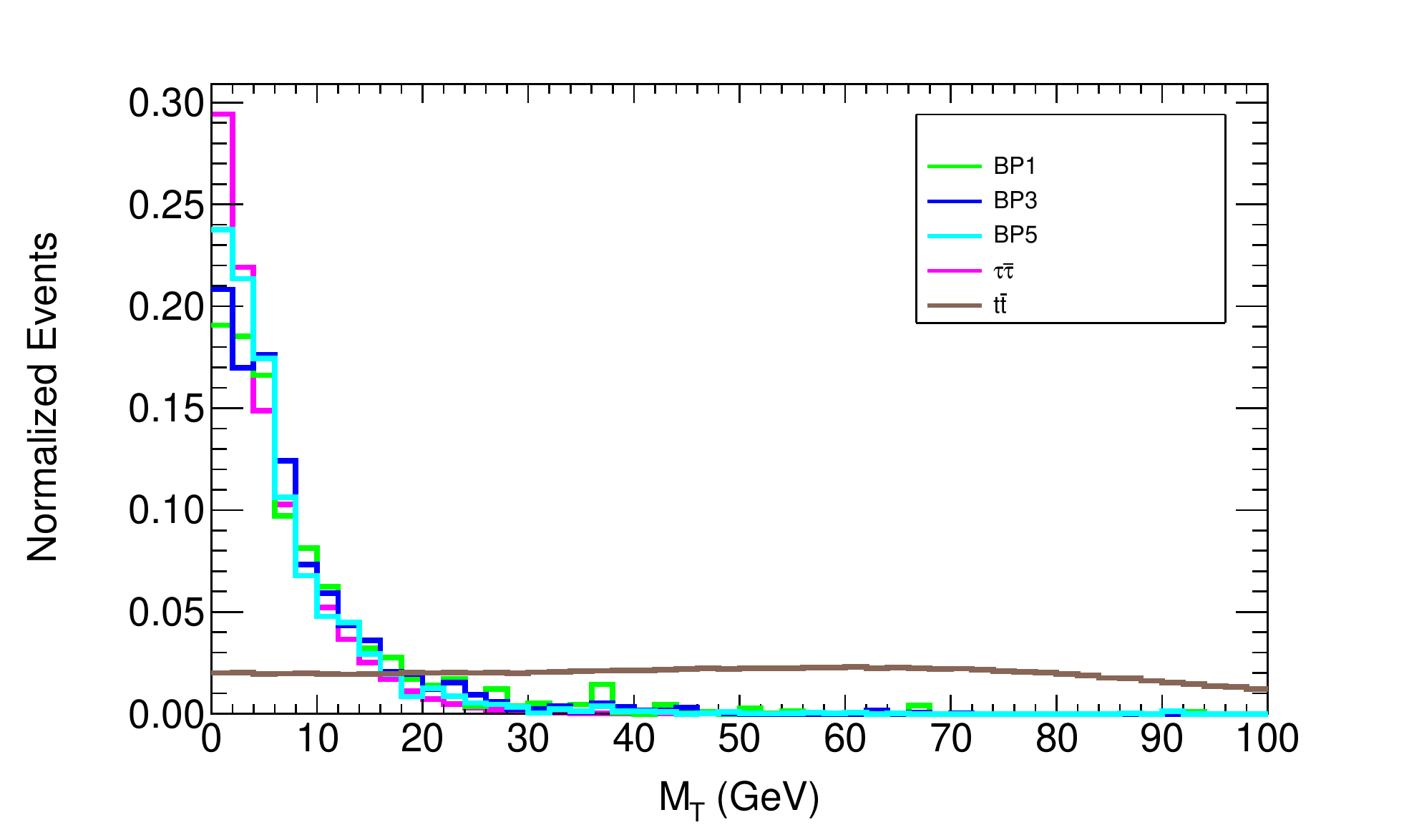}
 \caption{\it Distribution of collinear mass (left) and transverse mass (right) for signal and backgrounds.}
 \label{mcollinear_mtransverse}
 \end{figure}

 \begin{figure}[!hptb]
 	\centering
 	\includegraphics[width=7.2cm,height=6.0cm]{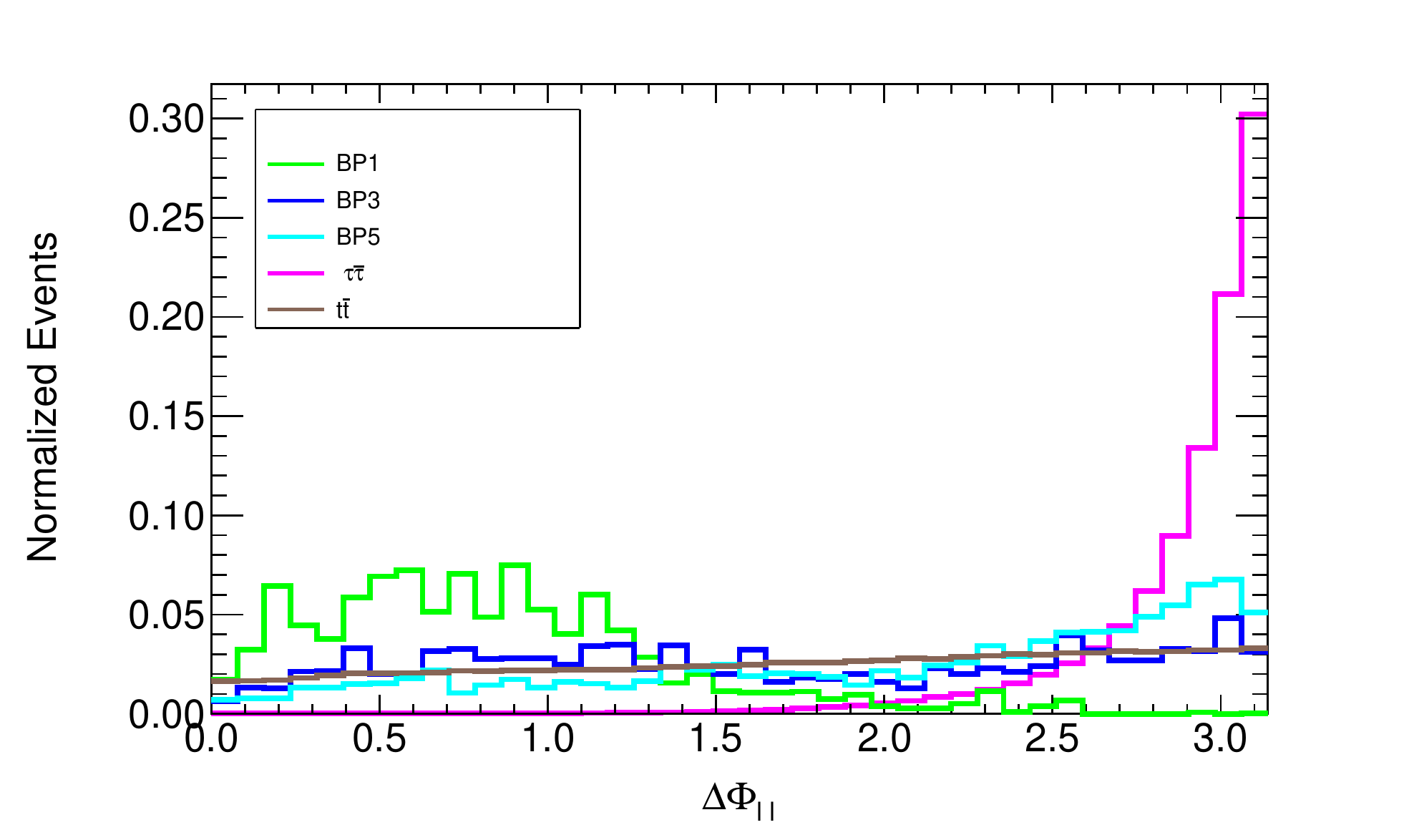}
 \caption{\it Distribution of $\Delta \phi$ between two leptons for signal and backgrounds.}
 \label{delpill}
 \end{figure}

           \item {\bf Missing transverse energy:} For the signal process, the only source of $\slashed{E_T}$ is the neutrino from the leptonic decay of $\tau$ in the final state which is again coming from the decay of a low mass pseudoscalar. Therefore, $\slashed{E_T}$ peaks at a lower value. For $\tau \tau$ background too, $\slashed{E_T}$ peaks appears at a lower value as the neutrinos, in that case, are almost back to back. Hence there is a significant overlap between the $\slashed{E_T}$ distribution from signal and $\tau \tau$ background. However, the $\slashed{E_t}$ produced in $t \bar t$ event peaks at a higher value. We present the $\slashed{E_T}$ distribution in Fig.~\ref{met_invll}(left).

   \item {\bf Invariant mass of the di-lepton pair:}  In Fig.~\ref{met_invll}(right) we show the invariant mass of the di-lepton system $M_{\ell\ell'}$. In the signal case, the leptons come from a low mass pseudoscalar, and therefore its distribution peaks at a much lower value, unlike the $\tau\tau$ and $t \bar t$ background.  $M_{\ell\ell'}$ plays a crucial role in reducing the $ee/\mu\mu$ background. The invariant mass for $ee/\mu\mu$ peaks at a $Z$-boson mass whereas the signal distribution peaks at a much lower value. By choosing a suitable cut on $M_{\ell\ell'}$, we can reduce this background. $M_{\ell\ell'}$ plays an important role to discriminate between the signal $\tau\tau$ background as well. In Table~\ref{tab:sig}, we show optimized cuts on $M_{\ell\ell'}$ for various benchmark points that we have applied to control the $\tau \tau$ background.

 \item {\bf The collinear mass:}  An important observable for our analysis is the collinear mass which is defined as follows:
       \begin{equation}
        M_{collinear} = \frac{M_{vis}}{\sqrt{x_{\tau_{vis}}}},
       \end{equation}
       where the visible momentum fraction of the $\tau$ decay products is,
       $x_{\tau_{vis}}=\frac{|\vec{p}_T^{\; \tau_{vis}}|}{|\vec{p}_T^{\; \tau_{vis}}|+|\vec{p}_T^{\; \nu}|}$,
       $\vec{p}_T^{\; \nu}=|\vec{\slashed{E}}_T| \hat{p}_T^{\; \tau_{vis}}$ and $M_{vis}$ is the visible mass of the $\tau - \ell$ system. 
The variable $M_{collinear}$ reconstructs the mass of the pseudoscalar from the $\slashed{E_T}$ and visible momenta. From Fig.~\ref{mcollinear_mtransverse} (left) it is evident that $M_{collinear}$ distribution shows a clear distinction between the signal and the $\tau \tau$ background. A suitable choice of cut on $M_{collinear}$ is imposed to reduce the $\tau\tau$ background (see Table~\ref{tab:sig}).

     \item {\bf The transverse mass:}  The next observable we considered is the transverse mass (Fig.~\ref{mcollinear_mtransverse} (right)) which is defined as
       \begin{equation}
          M_T(\ell) = \sqrt{2 p_T(\ell)\vec{\slashed{E}}_T (1 - \cos \Delta \phi_{\vec{\ell}-\vec{\slashed{E}}_T})}
       \end{equation}    
Here $ \Delta \phi_{\vec{\ell}-\vec{\slashed{E}}_T}$ is the azimuthal angle between the leading lepton and $\slashed{E}_T$. From Table~\ref{tab:sig} we can see that an optimized cut on $M_T$  has been applied to reduce the $t\bar{t}$ background.
           
   \item {\bf Angle between the lepton:}  The angle between two leptons $\Delta\phi_{\ell\ell'}$ is strongly correlated with the invariant mass of the di-lepton pair. Since for signal the invariant mass of the di-lepton pair peaks at a small value, the azimuthal angle between the two leptons $\Delta\phi_{\ell\ell'}$ shows a similar trend. On the contrary in the $\tau \tau$ background, the leptons are produced almost back to back and $\Delta\phi_{\ell\ell'}$ distribution peaks around $\pi$.
It is clear from Fig.~\ref{delpill} a suitable cut on this variable will help us enhance the signal over the background.

          \end{itemize}

 \begin{table}[ht!]
	\centering
	\scriptsize
	\begin{tabular}{|p{2.0cm}|c|c|c|c|c|c|}
		\cline{2-7}
		\multicolumn{1}{c|}{}& \multicolumn{6}{|c|}{Effective NLO cross-section after the cut(fb)}  \\ \cline{1-7}
		SM-background  
		& Preselection cuts &$\Delta\phi_{\ell\ell'} < 2.2$  & $M_{\ell\ell'}$ $ < 15$ GeV & $\slashed{E_T} < 15$ GeV  &  Mcollinear $> 10$ GeV & $M_T < 25$ GeV
		\\ \cline{1-7} 
                  $\tau\tau$ & 8582.75 & 132.089 & 0.21 & 0.089 & 0.052 & 0.052  \\ \cline{1-7} 
		$t\bar{t}$ leptonic & 25784.19  & 11.01 & 0.099 & 0.016 & 0.016 & 0.0016  \\ \hline \hline
			\multicolumn{1}{|c|}{Signal }  &\multicolumn{6}{|c|}{}   \\ \hline
		\multicolumn{1}{|c|}{BP1}& 0.574    & 0.446 & 0.164 & 0.157 & 0.150 & 0.148  \\ \hline \hline
		\multicolumn{1}{|c|}{BP2}& 0.619 & 0.374 & 0.0426 & 0.0398 & 0.0392 & 0.0385\\ \hline
		\end{tabular}
		\begin{tabular}{|p{2.0cm}|c|c|c|c|c|c|}
		\cline{2-7}
		\multicolumn{1}{c|}{}& \multicolumn{6}{|c|}{Effective NLO cross-section after the cut(fb)}  \\ \cline{1-7}
		SM-background  
		& Preselection cuts &$\Delta\phi_{\ell\ell'} < 2.2$  & $M_{\ell\ell'}$ $ < 20$ GeV & $\slashed{E_T} < 15$ GeV  &  Mcollinear $> 10$ GeV & $M_T < 25$ GeV
		\\ \cline{1-7} 
        $\tau\tau$ & 8582.75 & 132.089 & 0.576 & 0.278 & 0.239  & 0.239  \\ \cline{1-7} 
		$t\bar{t}$ leptonic & 25784.19  & 11.01 & 0.325 & 0.031 & 0.031 & 0.0016  \\ \hline \hline
		\multicolumn{1}{|c|}{BP3}& 0.402  & 0.196 &  0.0499 & 0.0451  & 0.0440 & 0.0438  \\ \hline 
        \multicolumn{1}{|c|}{BP4}& 0.236 &  0.0935 & 0.0118 & 0.001 &  & 0.0098 \\ \hline
  
	\end{tabular}

	\begin{tabular}{|p{2.0cm}|c|c|c|c|c|c|}
		\cline{2-7}
		\multicolumn{1}{c|}{}& \multicolumn{6}{|c|}{Effective NLO cross-section after the cut(fb)}  \\ \cline{1-7}
		SM-background  
		& Preselection cuts &$\Delta\phi_{\ell\ell'} < 2.2$  & $M_{\ell\ell'}$ $ < 30$ GeV & $\slashed{E_T} < 15$ GeV  &  Mcollinear $> 10$ GeV & $M_T < 25$ GeV
		\\ \cline{1-7} 
		
        $\tau\tau$ & 8582.75 & 132.089 & 12.79 & 9.15 & 9.11  & 8.81  \\ \cline{1-7} 
		$t\bar{t}$ leptonic & 25784.19  & 11.01 & 1.04 & 0.062 & 0.062 & 0.031  \\ \hline \hline
		\multicolumn{1}{|c|}{BP5}& 0.143  & 0.0439 & 0.0244  & 0.0191  & 0.0189 & 0.0186  \\ \hline 
		\end{tabular} 
		
          \begin{tabular}{|c|c|}
         \hline 
      Benchmark points &  Significance reach at 300 $fb^{-1}$ luminosity \\ \hline
      \hline 
  BP1 & 8.4 $\sigma$ \\ \hline
  BP2 & 2.6 $\sigma$\\ \hline
  BP3 & 1.5 $\sigma$\\ \hline
  BP4 & 0.3 $\sigma$\\ \hline
  BP5 & 0.1 $\sigma$\\ \hline
\hline
 \end{tabular}
	\caption{\it The cut-flow for signal and background and significance reach for our signal at 14 TeV LHC for 300 $fb^{-1}$ luminosity. }
	\label{tab:sig}
\end{table}

After applying optimized cuts on the relevant observables as listed in Table~\ref{tab:sig}, we obtain the signal significance for the benchmarks. The results are presented in Table~\ref{tab:sig} for 14 TeV, 300 $fb^{-1}$ luminosity. The significance~\cite{Cowan:2010js} has been calculated using the following formula.
\begin{equation}
{\cal{S}}=\sqrt{2[(S+B) Log(1+ \frac{S}{B}) -S]}
\end{equation}
where $S$ and $B$ denote the number of signal and background events after applying all the cuts respectively. We mention here that in order to take into account the 
next-to-leading-order (NLO) effects, we have multiplied the signal and background cross-sections with relevant {\it{k-factors}}. For signal, we take the $k$-factor of 2~\cite{Djouadi:2003jg} and for $t\bar{t}$ and $\tau\tau$ background, we use the $k$-factor to be 1.6~\cite{Sirunyan:2017uhy} and 1.15~\cite{Catani:2009sm} respectively. A comparison in terms of signal significance at the HL-LHC between the benchmarks from the ``wrong-sign" and ``right-sign''~\cite{Ghosh:2020tfq} is in order. As we mentioned earlier, in the ``wrong-sign'' case the cross-section can be higher than the ``right-sign'' case, for the same mass points. Therefore, the higher signal significance is achievable for the same benchmarks with lower luminosity. Moreover, it is possible to probe higher mass points in the ``wrong-sign'' case. However, we should mention that although it is possible to achieve a large cross-section in the ``wrong-sign'' case, it will be extremely difficult to probe beyond the mass scale that we considered due to the dominant contribution from the $\tau \tau$ and $ee/\mu\mu$ backgrounds.   


\subsection{Improved analysis with Artificial Neural Network (ANN)}

After the cut-based method, we analyze the di-lepton + $\slashed{E_T}$ final state with ANN~\cite{Teodorescu:2008zzb}. ANN has been extremely popular in the recent past~\cite{Hultqvist:1995ibm,Field:1996rw,Bakhet:2015uca,Dey:2019lyr,Lasocha:2020ctd} and it has been proved extremely effective to improve the results of cut-based analyses multi-fold~\cite{Dey:2019lyr,Dey:2020tfq,Bhowmik:2020spw}. In our present analysis where signal yield is poor, the signal and background separation becomes extremely crucial. In this regard, we have used ANN and calculated the maximum significance achievable at the HL-LHC with this technique. A python-based deep-learning library Keras~\cite{keras} has been used for ANN analysis.

Guided by our cut-based analysis we have chosen the input variables that yield large signal-background separation. The relevant observables and their definitions are listed in Table.~\ref{featurevar}. We have used these observables to train the network.

\begin{table}[htpb!]
\centering

\begin{tabular}{||c | c||} 
 \hline
 Variable & Definition \\ [0.5ex] 
 \hline\hline
 $p^{\ell_1}_{T}$ & Transverse momentum of the leading lepton \\ 
 $p^{\ell_2}_{T}$ & Transverse momentum of the sub-leading lepton \\
 $E^{miss}_{T}$ & Missing transverse energy \\
 $M_{\ell\ell'}$ & Invariant mass of the di-lepton pair \\
 $\Delta \phi_{\ell\ell'}$ & Azimuthal angle difference between the di-lepton pair \\ 
 $\Delta R_{\ell\ell'} $& $\Delta R$ separation between the di-lepton pair \\
 $M_{vis}$ & Visible mass of the di-lepton system \\
 $x_{vis}$ & Visible momentum fraction of the $\tau$ decay products \\
 $M_{collinear}$ & Collinear mass \\
 $M_T$ & Transverse mass \\
 $\Delta \phi_{\ell_1 \slashed{E_T}}$ & Azimuthal angle difference between the leading lepton and $\slashed{E_T}$ \\
 $\Delta \phi_{\ell_2 \slashed{E_T}}$ & Azimuthal angle difference between the sub-leading lepton and $\slashed{E_T}$ \\[1ex] 
 \hline
 \end{tabular}

 \caption{\it Feature variables used for training in the ANN analysis.}
  \label{featurevar}
\end{table}

We have used a network with four hidden layers with activation curve relu at all of them. The batch-size of 1000 is taken and the number of epochs per batch is 100. 80\% of the dataset has been used for training and 20\% for validation. It is crucial to avoid over-training of the data sample while doing the analysis. Over-training implies the training sample will yield extremely good accuracy but the validation or test sample will fail to achieve the same level of accuracy. We have explicitly checked that our network is not over-trained.

\begin{figure}[!hptb]
 	\centering
 	\includegraphics[width=9cm,height=7.0cm]{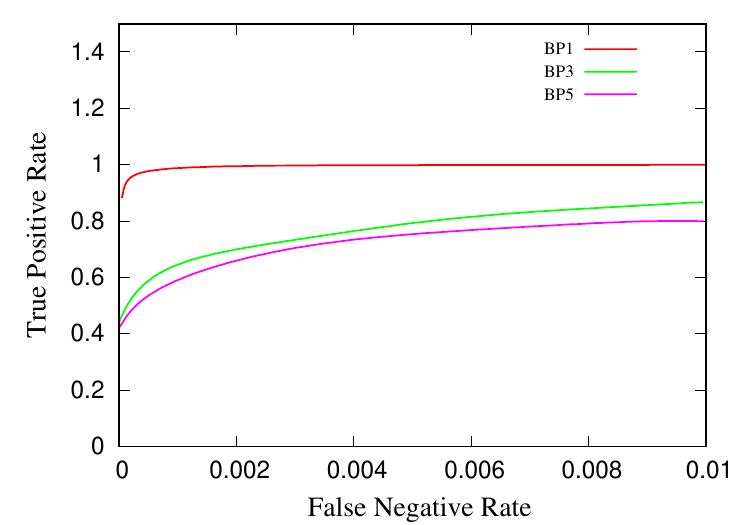}
 \caption{\it ROC curves for BP1, BP3 and BP5}
 \label{roc}
 \end{figure}

The variables $M_{\ell\ell'}$, $M_{collinear}$, $M_T$, $\Delta \phi_{\ell\ell'}$ and $\Delta R_{\ell\ell'} $ play the most important role in signal-background separation as was already clear from the cut-based analysis. However, there is a strong correlation between $\Delta R_{\ell\ell'} $, $\Delta \phi_{\ell\ell'}$ and $M_{\ell\ell'}$ which have been taken into account. We mention here to obtain a better performance from the network we have applied two basic cuts, namely $M_{\ell\ell'} < 30$ GeV and $M_{collinear} < 40$ GeV on signal and background events over and above the pre-selection. These cuts guide the network towards the signal region as can be seen from the distributions in the previous subsection and therefore enable better training. We obtain 99.9\%(BP1), 97.7\%(BP2), 95.4\%(BP3), 94.5\%(BP3),  and 89.0\%(BP5) accuracy, which indicates impressive signal-background separation. To avoid clumsiness, out of the five benchmark points we present in Fig.~\ref{roc}, the Receiver Operating Characteristic (ROC) curve for the BP1, BP3 and BP5 respectively.

The area under curve is 0.999(BP1), 0.998(BP2), 0.990(BP3), 0.988(BP4) and 0.987(BP5). We only show the part of the ROC curve which is relevant for our analysis. We scan over the points on the ROC curve and choose suitable points which yield the maximum signal significance for each benchmark. We present the signal significance ${\cal S}$ for all the signal benchmarks in Table.~\ref{significance_ann}.

\begin{table}[!hptb]
\begin{center}
\begin{tabular}{| c | c | c | }
\cline{1-2}
\multicolumn{1}{|c|}{BP} & \multicolumn{1}{|c|}{$
{\cal S}$ (cuts+ANN) at $300 fb^{-1}$}  \\ \cline{1-2}
BP1  & 12.6$\sigma$  \\ \cline{1-2}
BP2  & 8.8$\sigma$   \\ \cline{1-2}
BP3  & 5.4$\sigma$   \\ \cline{1-2}
BP4  & 2.5$\sigma$  \\ \hline \hline
 &  & \multicolumn{1}{|c|}{$
{\cal S}$ (cuts+ANN) at $3 ab^{-1}$} \\ \cline{1-3}
\multicolumn{1}{|c|}{BP5} & 0.8$\sigma$ & 2.5$\sigma$ \\ \hline
	
\end{tabular}

\caption{\it Signal significance for the benchmark points at 14 TeV LHC with cuts+ANN. }
\label{significance_ann}
\end{center}
\end{table}

Comparing the results of ANN in Table.~\ref{significance_ann} and that of the cut-based analysis in Table.~\ref{tab:sig} we can see that our analysis with ANN results in significant improvement for all the benchmarks.

\section{Conclusion}\label{conclusion}

In this work, we have considered generalized 2HDM with a Yukawa structure close to Type X 2HDM and have focused on the ``{\bf wrong-sign}" region of the parameter space. In this model, the non-standard scalar loops make a significant contribution to muon anomaly. On the other hand, the non-diagonal Yukawa couplings of this model naturally generate flavor violation in the leptonic sector. We have identified a parameter space with ``wrong-sign'' lepton-Yukawa coupling, which satisfies all the existing LFV constraints and simultaneously fits the most recently observed $g_{\mu}-2$ data. 

We then impose constraints coming from the requirement of perturbativity, unitarity and vacuum stability, measurement of oblique parameters, $B$-physics observables, and collider searches. We find that compared to the ``right-sign" region~\cite{Ghosh:2020tfq}, the ``wrong-sign" region gives rise to a different phenomenology which we explored in the present study. For example, unlike the ``right-sign" case, here one can have the lightest CP-even scalar as the 125-GeV Higgs, while the mass of the pseudoscalar is low, consistent with all the collider as well as theoretical constraints. Also, the non-standard CP-even and charged scalar masses can be much larger compared to the ``right-sign" case. In this work we have kept $m_H$ at 450 GeV and $m_{H^{\pm}}$ at 460 GeV. This choice in turn gives us more freedom to choose larger $\lambda_{bb}$ coupling and consequently makes allowance for much a larger production cross-section for low-mass pseudoscalar compared to the ``right-sign" case. 

We proceed next to the collider search for the flavor-violating decay of the low mass pseudoscalar to $\ell \tau \rightarrow \ell^+ \ell'^- + \slashed{E_T}$ final state, where $\tau$ decays leptonically and $\ell, \ell'= e, \mu$. First, we performed a cut-based analysis and find that with $300 fb^{-1}$ luminosity a mass range from 21 GeV to 26 GeV (BP1 and BP2) can be probed with significance $\gsim 2.5\sigma$ and for the BP3, BP4 and BP5 the significance is rather poor and even with 3$ab^{-1}$ luminosity one gets meager signal significance. We then perform an improved analysis using ANN and find that even with $300 fb^{-1}$ luminosity BP3 and BP4 can be probed with significance $\gsim 2.5\sigma$ and to probe BP5 with significance $ \gsim 2\sigma$ we need luminosity $\approx 3ab^{-1}$. We hereby point out that the ``wrong-sign'' region has a much better prospect compared to the ``right-sign" case~\cite{Ghosh:2020tfq} at the HL-LHC, in terms of detectability, since larger parameter space can be probed, with relatively lower luminosity in this scenario.


\section{Acknowledgement}

This work was supported by funding available from the Department of Atomic Energy, Government of India, for the Regional Centre for Accelerator-based Particle Physics (RECAPP), Harish-Chandra Research Institute.



\providecommand{\href}[2]{#2}\begingroup\raggedright\endgroup

\end{document}